\documentclass[journal]{IEEEtran}

\usepackage{amsmath,amssymb,amsfonts}
\usepackage{algorithmic}
\usepackage{graphicx}
\usepackage{textcomp}
\usepackage{xcolor}
\usepackage{listings}
\usepackage{multirow}
\usepackage{multicol}
\usepackage{url}
\usepackage{float}

\usepackage[numbers]{natbib}

\usepackage{pgfplots}
\usepgfplotslibrary{dateplot}
\pgfplotsset{compat=1.12}
\usetikzlibrary{fillbetween}
\usepackage{filecontents}

\ifCLASSOPTIONcompsoc
    \usepackage[caption=false, font=normalsize, labelfont=sf, textfont=sf]{subfig}
\else
\usepackage[caption=false, font=footnotesize]{subfig}
\fi

\def\BibTeX{{\rm B\kern-.05em{\sc i\kern-.025em b}\kern-.08em
    T\kern-.1667em\lower.7ex\hbox{E}\kern-.125emX}}
    
\usepackage{color,soul}
\soulregister\cite7
\soulregister\ref7
\renewcommand\hl[1]{#1}

\begin{document}

\title{Academic Source Code Plagiarism Detection by Measuring Program Behavioural Similarity}

\author{
\IEEEauthorblockN{Hayden Cheers, Yuqing Lin and Shamus P. Smith}

\IEEEauthorblockA{School of Electrical Engineering \& Computing\\
The University of Newcastle, Callaghan, NSW, Australia, 2308\\
Email: \{hayden.cheers, yuqing.lin, shamus.smith\}@newcastle.edu.au} 
}

\maketitle

\begin{abstract}
Source code plagiarism is a long-standing issue in tertiary computer science education. Many source code plagiarism detection tools have been proposed to aid in the detection of source code plagiarism. However, existing detection tools are not robust to pervasive plagiarism-hiding transformations, and as a result can be inaccurate in the detection of plagiarised source code. This article presents BPlag, a behavioural approach to source code plagiarism detection. BPlag is designed to be both robust to pervasive plagiarism-hiding transformations, and accurate in the detection of plagiarised source code. Greater robustness and accuracy is afforded by analysing the behaviour of a program, as behaviour is perceived to be the least susceptible aspect of a program impacted upon by plagiarism-hiding transformations. BPlag applies symbolic execution to analyse execution behaviour and represent a program in a novel graph-based format. Plagiarism is then detected by comparing these graphs and evaluating similarity scores. BPlag is evaluated for robustness, accuracy and efficiency against 5 commonly used source code plagiarism detection tools. It is then shown that BPlag is more robust to plagiarism-hiding transformations and more accurate in the detection of plagiarised source code, but is less efficient than compared tools.
\end{abstract}


\section{Introduction}

Plagiarism is a long-standing issue in academic institutions. Studies have indicated between 50\% to 79\% of undergraduate students will plagiarise at least once during their academic careers \cite{yeo2007, sraka2009, curtis2011, pierce2017}. With such a high rate, it is highly likely that an academic will have to at one point assess a suspected case of plagiarism. In computer science courses, plagiarism is commonly encountered as source code plagiarism. \textit{Source code plagiarism} occurs when one student appropriates the source code of another and proceeds to submit it as their own work. Subsequently, source code plagiarism can be suspected when either: one assignment shares a large subset of code with another \cite{cosma2008} (e.g. via copy and paste); or one assignment is a complete copy of another \cite{parker1989}. However, plagiarism can be a difficult and time-consuming task to identify \cite{joy1999}, often requiring a large effort on the part of academics to review and assess assignment submissions for plagiarism.

To aid in the detection of source code plagiarism, many \textit{Source Code Plagiarism Detection Tools} (SCPDTs) \cite{martins2014,novak2019} have been proposed. A SCPDT analyses a pair of assignment submissions to identify indications of plagiarism. This is typically by evaluating the similarity of submission pairs by measuring specific aspects of source code. The measured aspects of source code are SCPDT-dependent, however, in most SCPDTs, the similarity score indicates what percentage of one submission can be found in another. Subsequently, a high similarity score implies plagiarism has occurred (due to a large overlap of source code), a mid-range similarity score may imply students have collaborated on an assignment (a form of academic misconduct), while a low similarity implies plagiarism has not occurred. 

A plagiarising student may attempt to hide the act of plagiarism to evade detection. This is by applying source code transformations to a misappropriated assignment, causing it to appear superficially distinct. Such transformations are typically cosmetic by changing the appearance and structure of the source code \cite{parker1989, joy1999, faidhi1987}, but retaining the original behaviour of the plagiarised assignment. In this work, such transformations are referred to as \textit{plagiarism-hiding transformations}. Subsequently, it is desirable for a SCPDT to be:
\begin{enumerate}
    \item Robust to plagiarism-hiding transformations.
    \item Accurate detecting plagiarised and transformed works.
\end{enumerate}

Robustness and accuracy are related but distinct qualities of a SCPDT. A SCPDT is robust when it can accommodate for plagiarism-hiding transformations. This will result in a SCPDT reporting a high similarity between a plagiarised assignment with applied source code transformations and its source. Similarly, a SCPDT is accurate when it measures a high level of similarity between a plagiarised assignment and its source to imply plagiarism is present; while also measuring a low similarity between unrelated works, implying that plagiarism is not present. 

A SCPDT can gain robustness to plagiarism-hiding transformations by ignoring aspects of source code that plagiarisers commonly modify (e.g. ignoring comments or identifiers in source code \cite{prechelt2002}). However, if a SCPDT ignores too many aspects of a program, the SCPDT may measure greater similarity between unrelated works. This can result in the SCPDT being inaccurate by indicating unrelated works as being plagiarised (a false positive). Conversely, if a SCPDT is not robust to plagiarism-hiding transformations, and analyses aspects of source code that change due to plagiarism-hiding transformations, a SCPDT may become too reserved in measuring similarity. This can result in the tool failing to measure a high enough similarity to indicate plagiarism (a false negative). Therefore, a SCPDT should be both robust and accurate in the detection of plagiarism.


Although existing SCPDTs are largely robust and accurate in the presence of plagiarism-hiding transformations, they are vulnerable to \textit{pervasive} plagiarism-hiding transformations \cite{cheers2020}. \textit{Pervasive plagiarism-hiding transformations} represent extreme cases of plagiarism-hiding transformations, where a plagiariser has applied source code transformations entirely throughout a body of source code, transforming it such that it bares little cosmetic or structural similarity. Such transformations can greatly impact upon the ability of SCPDTs to measure similarity. Subsequently, when a plagiarised work is pervasively transformed, SCPDTs can report a low enough similarity that it does not raise suspicion of plagiarism \cite{cheers2020}. 

To address this issue, this article presents a novel approach to source code plagiarism detection (SCPD) titled \textit{BPlag}. \hl{BPlag identifies indications of plagiarism through the analysis of behavioural similarity between two programs as expressed through their source codes. At no point is the source code of two programs directly compared to measure similarity. Instead, the execution behavior of two programs (in terms of how data is used and transformed, and how the program interacts with the execution environment) is used to derive a measure of behavioural similarity. BPlag applies symbolic execution to record a programs execution behaviour.} The behaviour of a program is subsequently represented in a novel graph-based format that is used to evaluate similarity of two programs\hl{, and subsequently identify indications of plagiarism}. Using this method, BPlag is shown to be both robust to plagiarism-hiding transformations and accurate in the detection of pervasively transformed plagiarised assignments. BPlag is available from \cite{bplag}.

The remainder of this article is structured as followed. Section II presents background on approaches to measuring program similarity and plagiarism-hiding transformations. Section III discusses the use of behavioural similarity for SCPD, and its expected benefits. Section IV presents the design and justification of BPlag. Section V presents a comprehensive evaluation of BPlag in comparison to five commonly used SCPDTs. Sections VI and VII discuss the significance of BPlag and compare it to related works. Lastly, section VIII concludes this article and identifies future directions of work. 

\section{Background}

\subsection{Measuring Program Similarity}

Approaches to measuring the similarity of two programs can be broadly classified by what aspects of the programs are being compared. While there are many unique and novel approaches for measuring similarity, in this work they are generalised as being either: structural, semantic or behavioural. Note, other works sometimes term semantic approaches as measuring behavioural similarity (e.g. \cite{sihan2016}). In this article, a behavioural approach for measuring similarity is considered to require dynamic (i.e. runtime) analysis of a program.

Structural approaches measure similarity by identifying common structures in source code. In its most basic form, structural similarity can be measured with textual strings. This is by applying techniques such as string edit distance or string alignment to measure the similarity of source code \cite{joy1999,pikeunknown,gitchell1999a,rani2018}. However, it is more common to see structural similarity measured with the comparison of lexical token sequences, representing the structure of the source code in terms of important lexical elements. Subsequently, structural similarity can be measured with token edit distance or tiling-based approaches \cite{joy1999, prechelt2002, schleimer2003, ahtiainen2006, grune1989, anzai2019}. Other approaches measure the structural similarity of parse trees or abstract syntax trees, representing the source code within the grammar of a programming language \cite{li2010,zhao2015,fu2017}.  

Semantic approaches measure similarity through the meaning of source code. This is through semantic analysis, that analyses source code to extract information not expressed through the grammar of a programming language. Semantic approaches typically analyse a program through program dependence graphs \cite{ferrante1987}. The program dependence graph identifies the relations between terms within a procedure or method. Subsequently, the similarity of these graphs can be calculated (e.g. with graph edit distance or sub-graph embedding) to identify indications of plagiarism (e.g. \cite{liu2006, chen2010, chae2013}). While other methods include applying latent semantic analysis to identify similarly referenced terms \cite{cosma2012} or identifying similar call graph structures \cite{prado2018}. 

Behavioural approaches analyse the runtime behaviour of a program. There are a diverse range of techniques applied to identify behavioural similarity. This can be by analysing the functional equivalence of a program \cite{sihan2016,bertran2005}, the use of data at runtime \cite{jhi2011}, identifying similar interactions with the execution environment \cite{anjali2015}, or identifying similar program logic \cite{zhang2014,luo2017}. Such approaches are based on the assumption that the behaviour of a program is a uniquely identifying feature, and that similar behaviours indicate similar programs. 

\subsection{Plagiarism-Hiding Transformations}

Plagiarism-hiding transformations are very broad. There is are countless methods that can be used to modify a program such that plagiarism can go unnoticed. Joy \& Luck \cite{joy1999} provided examples of source code transformations used by students to hide plagiarism. These transformations are grouped as lexical (i.e. cosmetic) and structural changes. For example, lexical changes modify identifiers or the formatting of source code; while structural changes modify statements and expressions in source code. A commonly referenced taxonomy of source code transformations applied to hide plagiarism was presented by Faidhi and Robinson \cite{faidhi1987}. This taxonomy categorises transformations into 6 levels:
\begin{enumerate}
    \item[L1)] Changes to comments and indentation.
    \item[L2)] Changes to identifiers.
    \item[L3)] Changes in declarations (e.g. declaring extra constants, changing the order of functions and variables).
    \item[L4)] Modifying functions (e.g. modifying signature, merging and creating new functions).
    \item[L5)] Changing program statements to semantic equivalents (e.g. for to while, if to switch).
    \item[L6)] Changes in decision logic and modifying expressions.
\end{enumerate}

In this taxonomy, each level of transformation includes the transformations of the previous. For example, L6 transformations include all L1 to L5 transformations. Each level of transformation implies a greater difficulty in application to preserve the original program semantics and hence requires a greater understanding of the program. However, it does not imply that higher levels of obfuscation require more effort, nor that it has a greater impact on reducing similarity. The lower levels are more typical of novice programmers, and higher levels are more representative of skilled programmers. 

Notably, the majority of these transformations do not modify the execution behaviour of a program. This is as a plagiariser is commonly somebody who is either: inept, lazy, time-poor \cite{joy1999, sheard2003, simon2019}. In order to correctly modify the behaviour of a program, the plagiariser would have to take the time to understand the program and be skilled enough not to break it in the process. Hence, typically the original behaviour of a plagiarised program remains largely unchanged. 

\section{Behaviour for Academic Source Code Plagiarism Detection}

\hl{
When analysing program behaviour for identifying indications of plagiarism, it raises the question of what aspects of behaviour are most appropriate for measuring similarity. Zhang et. al. \cite{zhang2014} describe program behaviour at a high level as \textit{``input, output and the computation used to achieve the input-output mapping''}. This serves as a basis for use in source code plagiarism detection, however the nature of assignment submissions need to be taken into account.

All assignment submissions implementing the same assessment task are all inherently behaviourally similar. This is as they implement the same assignment specification, meaning it can be expected that all assignment submissions for the same assessment task will accept the same inputs, and if correctly implemented, produce the same outputs. The mapping of inputs to outputs as a method of measuring similarity is termed functional equivalence, i.e. programs $A$ and $B$ are functionally equivalent if given inputs $I$, they both produce outputs $O$. Therefore, comparing the behaviour of a program with functional equivalence is not suitable in source code plagiarism detection, as it would result in all correctly implemented submissions being detected as identical.

By discounting input and output, the \textit{``computation used to achieve the input-output''} needs to be considered in how it can be represented and applied to measure similarity. To derive a behavioural representation of the computation of a program, its nature is considered from a high level object-oriented perspective, and contrasted with a low-level procedural perspective. This is exemplified using Java, as Java SCPD is the focus of this work. From an object-oriented perspective, a program is a series of objects communicating. Each object has state (data) and relations that are transformed through the execution of a program. However, from a procedural perspective, a program is fundamentally a series of primary operations on primitive values, causing it to change state. Both perspectives share the calling of services from the execution environment through method calls, and passing and receiving data in the same process. 

Java is an object-oriented language, and as such, there is a wealth of knowledge expressed in the object-oriented design of the program. Such knowledge indicates how a problem (i.e. an assignment specification) has been modelled by a programmer and is subsequently implemented. This is expressed in how objects are composed, interact and the logic defined within them. Therefore, the object-oriented nature of the program needs to be considered by representing the composition of objects and the relations between them. To describe the procedural nature a program, it is considered in terms of binary instructions as Java bytecode. The Java byte code instruction set}\footnote{https://docs.oracle.com/javase/specs/jvms/se8/html/jvms-6.html}\hl{ describes three categories of instruction: data-oriented (create/read/store values), arithmetic-oriented (mutate values with mathematical or logical operations), and branching-oriented (conditional execution and method calls). These categories of instructions allow for describing the behaviour of a program through three important aspects: data, the transformation of data, and the invocation of methods. These three aspects essentially describe a program from a \textit{procedural} perspective. If these perspectives are combined, the behaviour of a program, while also including a high-level object-oriented design can be described as three aspects:}
\begin{enumerate}
    \item Data and its relations.
    \item The transformation of data through arithmetic operations.
    \item Interactions with the execution environment.
\end{enumerate}

\hl{Fundamentally, a Java program is no more than these three aspects. Hence, they are an important method of describing an approximation of the computation of a Java program. Using this model, two behaviourally-equivalent programs have the same design, implement the same processes, and execute the same operations at runtime. This model will be used for analysis and measurement of behavioural similarity of Java programs, applied for SCPD.}
 
\hl{However, this model can only be termed to provide an `approximate' representation of the execution behaviour of a program. The term `approximate' is used as there are other aspects of execution behaviour that have been analysed in prior works, but not included here. For example prior works have analysed: input-output relations \cite{zhang2014}, executed instruction sequences (i.e. execution traces) \cite{tian2015}, stack usage patterns \cite{park2015}, heap structures \cite{chan2012}, and API method call sequences \cite{anjali2015,wang2009}. However, these three analysed aspects are sufficient for representing behaviour for source code plagiarism detection. This will be demonstrated through the evaluation of the proposed SCPDT, BPlag, that implements this model (section \ref{s:ev}).}

\subsection{Behaviour and it's Robustness to Plagiarism-Hiding Transformations}

Analysing program behaviour is expected to be most beneficial in the detection of plagiarism with pervasively applied structural transformations. Structural transformations are equivalent to L3+ transformations from Faidhi and Robinson's \cite{faidhi1987} taxonomy. Such transformations modify the appearance and structure of source code, but not behaviour. Applying pervasive structural transformations is expected to be performed by ``skilled but time-poor'' plagiarisers. This is a student who has the skills to implement an assessment item, but does not have the time or motivation to do so. 

It is conceivable that such a student has the skills to pervasively transform source code such it no longer warrants suspicion of plagiarism. Cheers, Lin and Smith \cite{cheers2020} noted that with the use of modern integrated development environments, many automated source code transformations can be applied to a misappropriated assignment in order to transform it such that currently available SCPDTs would not measure a high enough similarity to warrant suspicion. Such transformations took approximately 1-2 hours to apply, being much less time than a typical major programming assignment. For example, consider a major programming assignment to be at least 10 hours work. 

While it is easy to transform the structure of a program, it is argued that it is much more difficult to transform the behaviour of a program to evade detection. Using the utilised aspects of behaviour, in order to change the execution behaviour of a program into an unrecognisable form, a plagiariser would have to gain intimate knowledge of a misappropriated work and reverse engineer it into a functionally-equivalent but a behaviourally-distinct form. This means a plagiariser would have to transform the implementation of a programs execution behaviour such that it uses a different data model, different process of transforming data, and different interactions with the execution environment; while retaining the original input-output relations. To do so, the plagiariser would effectively have to rewrite significant portions of the mis-appropriated program. When considering the motivations of a plagiariser, it is unlikely they would spend so much time to hide their plagiarism.

\section{BPlag}

By combining both high-level and low-level perspectives of a program, the behaviour of a program is expressed through the three identified aspects. This allows for representing an approximation of program behaviour as a reflection of a programs conceptual design and implementation. Subsequently, this allows for a behavioural approach to SCPD that can differentiate between programs that are behaviourally similar due to implementing the same specification, and those that are behaviourally similar due to plagiarism; in a manner that is both robust and accurate.

BPlag models the behaviour of a program using the three identified aspects: data and its relations; the transformation of data through arithmetic operations; and interactions with the execution environment. \textit{Data} is considered to be a value in memory, that may be a primitive value or complex object. Complex objects have \textit{Relations}, indicating concepts such as aggregation and ownership. \textit{Transformations} occur when primary operations are executed upon primitive values (through language operators). \textit{Interactions} with the execution environment are calls to library methods. As the implementation of BPlag is for Java SCPD, the execution environment is the Java Virtual Machine (JVM), and library methods are defined in .jar files.

\begin{figure}
    \centering
    \includegraphics[width=\linewidth]{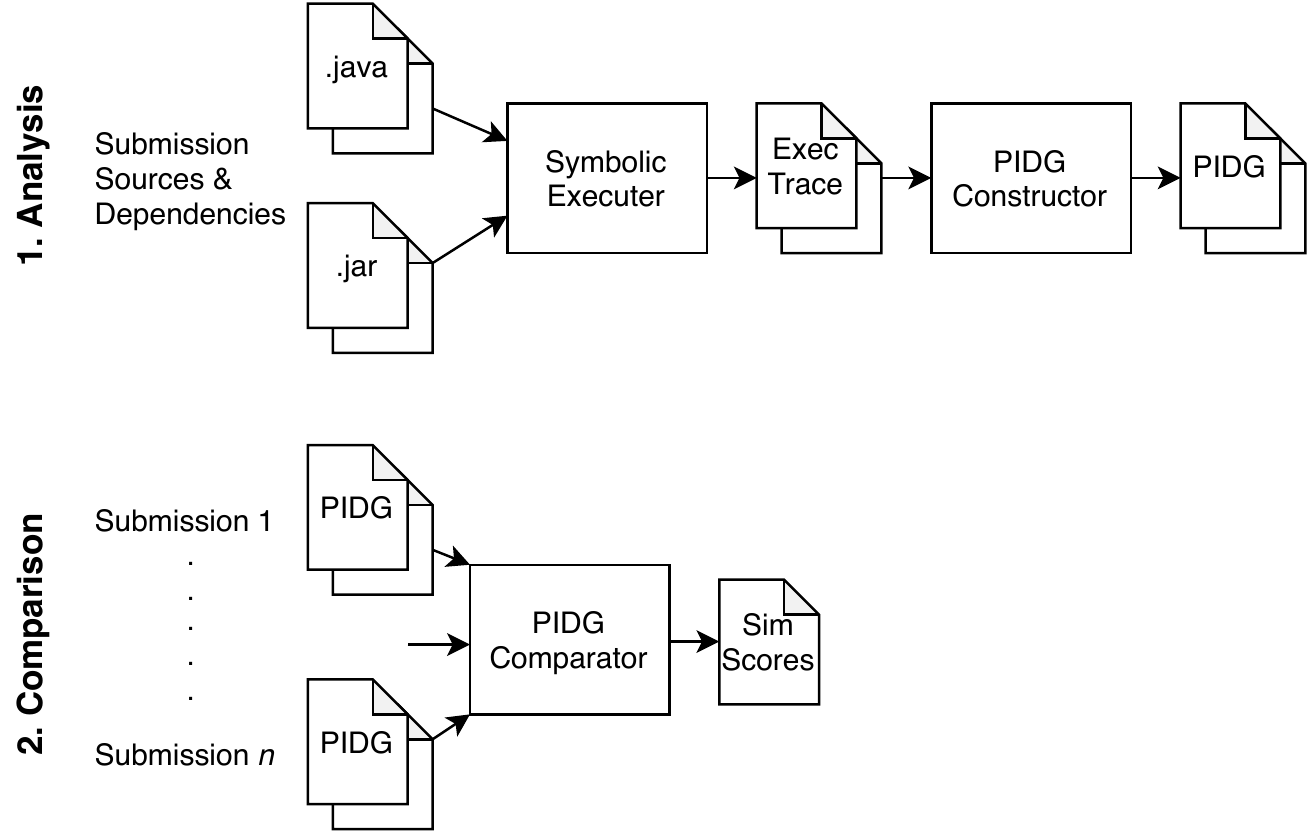}
    \caption{Overview of BPlag. Firstly, assignment submissions are analysed and transformed to represent program behaviour. Secondly, the submissions are compared for behavioural similarity, reporting a list of similarity scores.}
    \label{fig:bplag_overview}
\end{figure}

The design of BPlag is divided into two phases. Firstly, assignment submissions are analysed to derive an approximate behavioural representation. Secondly, the behavioural representation of all programs are analysed for similarity. Fig. \ref{fig:bplag_overview} presents an overview of this process. This is decomposed into three main components: 
\begin{enumerate}
    \item A symbolic execution tool to record and extract a program's execution behaviour.
    \item A component to construct the behavioural representation of a program as a set of Program Interaction Dependency Graphs \cite{cheers2020-icsim} (PIDG).
    \item A component to evaluate analyse the similarity of two programs by comparing PIDGs.
\end{enumerate}

The behaviour of a program is extracted through symbolic execution \cite{baldoni2018}. Symbolic execution is a program testing and validation technique that simulates the execution of a program. Typically it is applied to determine what set of inputs cause specific code to be executed in a program. BPlag applies symbolic execution to analyse and record the execution behaviour of a program. The behaviour of a program is extracted as a set of execution traces, recording important execution events. 

The execution traces extracted from a program are used to construct a set of PIDGs. A single PIDG represents the behaviour of one execution of a program, identifying the utilised data and its relations, how the data is transformed and mutated, as well as interactions with the execution environment as API method calls. A set of PIDGs are used to represent the entire behaviour of a program across all possible executions. The behavioural similarity of two programs can then be evaluated by comparing the sets of PIDGs for similarity. This is by identifying similar data with similar relations, that are transformed by similar operations, and are used in similar interactions with the execution environment.

The current implementation of BPlag is designed as a Java SCPDT. However, the fundamental principals can be applied to any similar high-level programming languages (e.g. C\#, Python) by utilising an appropriate symbolic execution tool. The following sub-sections describe each component.

\subsection{Symbolic Executor}

The symbolic executor component simulates the execution of a program. It is the core component of BPlag that enables the recording of execution behaviour. Symbolic execution is applied in BPlag as it is perceived to provide greater coverage in the analysis of execution behaviour. To derive a comprehensive behavioural representation of a program, all possible execution paths of a program need to be explored. If concrete execution (i.e. `normal' execution) was applied to analyse behaviour, it would require providing a large set of test inputs to cause a program to execute multiple different execution paths to record behaviour under different inputs sets. This was deemed infeasible for a SCPDT. However, recording all potential execution paths is easily afforded with symbolic execution. This is achieved by providing symbolic values as input to the program, causing it to execute all possible execution paths. Hence, with symbolic execution, BPlag can analyse any valid Java program without user-provided test inputs.

\begin{figure}
    \centering
    \includegraphics[width=\linewidth]{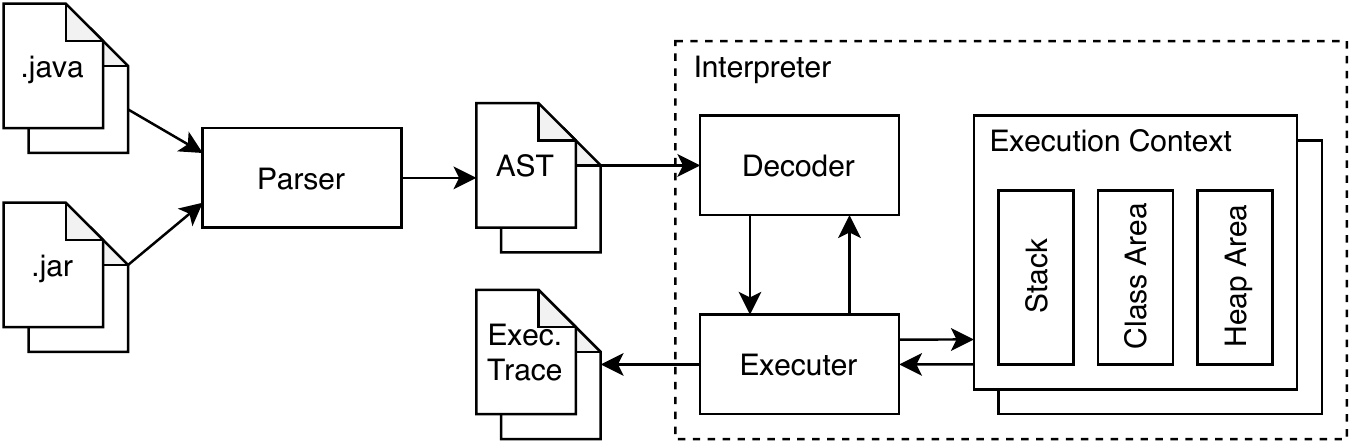}
    \caption{Symbolic executor design. The symbolic executor is modelled after the JVM, however is implemented as a tree-walk interpreter.}
    \label{fig:symbolicexecutor}
\end{figure}

\hl{The purpose-built symbolic executor is designed for recording the three aspects of behavioural similarity analysed by BPlag. As such it enables fine-grained analysis of a program's execution behaviour, and is optimised for the requirements of a SCPDT.} Fig. \ref{fig:symbolicexecutor} presents the design of the symbolic executor. The symbolic executor is modelled after the Java 8 virtual machine. However, instead of accepting Java bytecode as input (as per existing symbolic execution tools), it accepts Abstract Syntax Trees (AST) that are interpreted. ASTs are parsed using the Eclipse JDT toolkit\footnote{https://www.eclipse.org/jdt/} and interpreted with a simple tree-walk. Technically, this component is architected as a `symbolic source code interpreter'. Hence, it would be more appropriate to term it a `symbolic interpreter' (as a Java symbolic executor would execute Java bytecode), however for simplicity, it is referred to as an executor. 

The symbolic executor supports exploring all potential execution paths defined by a program. The executor encapsulates the state of a single execution path in an execution context. The execution context maintains a stack of execution frames, each containing: the instructions being executed, operands the instructions execute upon, a reference to the class area (also referred to as the meta-space), and the heap area (where object types are stored). When the executor encounters a branching statement (i.e. if or switch), it will clone the current context (containing the execution frame stack, class area and heap area), enabling a fork to explore all subsequent paths independently. 

The symbolic executor only supports executing source code, it does not execute Java bytecode. Subsequently, all interactions with the execution environment (as API method calls) are stubbed. For example, when calling a method defined in source code, it is executed as appropriate; however, calling an API method call (defined in a .jar file) results in a stubbed interaction. Subsequently, when an API method is invoked, the invocation is recorded and symbolic data is returned as appropriate. This was deemed appropriate as the approach is only concerned with representing the behaviour of the user-written program expressed through source code. The behaviour of an API method is not considered important in this context.

\begin{figure}
    \centering
    \includegraphics[scale=0.75]{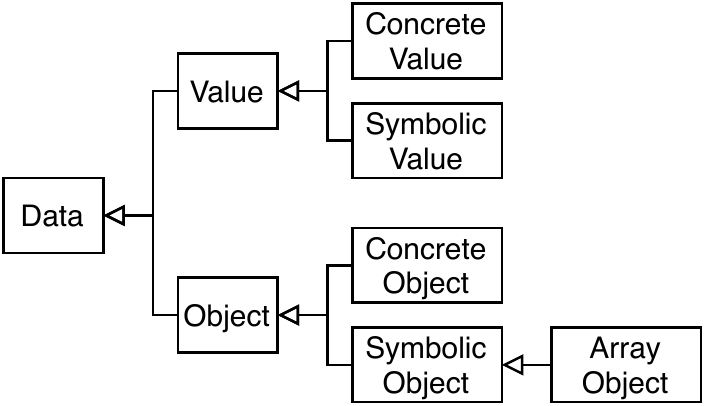}
    \caption{Types of data utilised by the symbolic executor component.}
    \label{fig:data_hierachy}
\end{figure}

Data (see Fig. \ref{fig:data_hierachy}) is modelled by the executor as either values or objects. Values are one of the eight Java primitive types (byte, short, int, long, float, double, char, boolean). Objects (and arrays) are java class-types. Both values and objects come in `symbolic' and `concrete' variants. A concrete value is well-known at runtime (i.e. can be resolved to an exact value); similarity, concrete objects are of a well-known type with a well-defined set of fields. Conversely, a symbolic value is represented as a symbol that has mathematical constraints upon it; while symbolic objects have the potential to have fields dynamically `created' and assigned to. Arrays are special cases of objects and are considered symbolic as any interaction can result in storing a value at an index represented by a symbolic value. As a generalisation, concrete data is created directly in user-defined code and have well-known values; while symbolic data is created by interactions with the execution environment.

\begin{table}[]
    \centering
    \caption{Recorded Execution Trace Events}
    \begin{tabular}{|p{2.5cm}|p{5cm}|}
        \hline
        Event & Recorded Event Parameters \\
        \hline
        \hline
        \multirow{4}{*}{API method call} & Method signature \\
         & Scope (Object, instance methods only) \\
         & Parameters (array of Data) \\
         & Result (Data, non-void methods only) \\
        \hline 
        \multirow{4}{2.5cm}{Field assignment \& retrieval} & Static type (static field only) \\
         & Scope (Object, instance fields only) \\
         & Field name \\
         & Assigned/retrieved data (Data) \\
        \hline
        \multirow{3}{2.5cm}{Array index assignment \& retrieval} & Scope (Array) \\
         & Index (Value) \\
         & Assigned/retrieved value (Data) \\
        \hline 
        \multirow{3}{2.5cm}{Primary operation} & Operator (arithmetic, comparison or logical) \\
         & Operands (Data, labelled LHS \& RHS) \\ 
         & Result (Value) \\
        \hline 
        \multirow{2}{2.5cm}{String concatenation} & Operand (array of Data, of type String) \\
         & Result (Data, of type String) \\
        \hline 
        \multirow{2}{2.5cm}{Stringification of values} & Operand (Data) \\
         & Result (Object, of type String) \\
        \hline 
        \multirow{2}{2.5cm}{Asserted constraints on symbolic values} & Assertion (Value, as symbolic boolean) \\
         & Truth (true/false value) \\
        \hline 
    \end{tabular}
    \label{tab:exec_trace_events}
\end{table}

The executor records important events that reflect a program's behaviour, listed in table \ref{tab:exec_trace_events}. These events form the execution trace, identifying what data is used, how it is transformed, and how the program interacts with the execution environment. \hl{Most recorded events correspond to language-level statements, and occur during both concrete and symbolic execution. This allows for recording what data is used during execution, the relations between data, how it is transformed, and any interactions with the execution environment. The only unique events recorded by the symbolic executor are \textit{``asserted constraints on symbolic values''}. A symbolic value by definition does not have a well-known value. However, if it is used in a conditional expression (e.g. in an \textit{if} branch), the executor is required to assert that it is either true or false, and continue along the appropriate execution path. In the current implementation of the symbolic executor, both cases are explored independently (i.e. with a conditional value asserted as true and false). However, such assertions are not used by BPlag to measure similarity, as conditional constraints are not an analysed aspect of behaviour.}

\subsubsection{Comparison with Existing Symbolic Execution Tools}

\hl{
There exist many Java symbolic execution tools that could have been applied in BPlag. For example, JPF\footnote{https://babelfish.arc.nasa.gov/trac/jpf/wiki/projects/jpf-symbc}, JDart\footnote{https://github.com/psycopaths/jdart}, and JBSE\footnote{http://pietrobraione.github.io/jbse/}. However, a custom-built symbolic execution tools is used due to three main quality concerns:}
\begin{enumerate}
    \item Computational complexity of existing tools
    \item Impacts of compilation on execution behaviour
    \item Quality of student assignments for analysis
\end{enumerate}

\hl{
Firstly, existing symbolic execution tools are designed for applying symbolic execution to program testing. In particular, they seek to identify the constraints on inputs that result in certain execution paths to aid software quality activities such as software testing or identifying bugs in code \cite{baldoni2018}. Analysing the constraints placed upon inputs requires the use of a constraint solver that can determine if encountered path conditions are mutually compatible/exclusive. Solving constraints is a computationally intensive process, and is not required by BPlag. Furthermore, existing symbolic execution tools execute \textit{all} Java bytecode, even that provided as part of system libraries. BPlag is not concerned with the behaviour of system libraries, hence by mocking their interaction it lowers the amount of computation. Hence, a custom built tool that omits constraint solving and library bytecode execution was deemed to provide greater efficiency.

Secondly, during the compilation pipeline, compilers are known to optimise source code. While the standard Java compiler is known not to optimise bytecode, it was observed that it may rewrite and transform conditional code. In order to eliminate any risks associated with using compiled code, it was deemed that a source code interpreter would be appropriate. Furthermore, as BPlag is a SCPDT, it was desirable to record program behaviour as \textit{expressed through} source code, and not as bytecode.

Thirdly, the quality of student code is known to vary. It is not uncommon to see assignment submissions with syntactic or semantic errors in them. If an existing symbolic execution tool was applied, it would stop BPlag from analysing such submissions; as it would require analysing only compilable assignments. When using a source code interpreter, there is the potential for submissions with syntactic or semantic errors to be partially-evaluated. In a best-case scenario this would result in some execution paths terminating early, while in a worst case scenario this would prohibit assignments from being compared. Furthermore, as most existing SCPDTs only require submissions to be parseable, this was another desired quality of BPlag.}

\hl{The symbolic executor also makes some assumptions regarding the execution of any program to accommodate for the quality of student code. This is for largely for stability, and to bring assurances that any analysed program will terminate. Firstly, the execution of any loop (i.e. for, enhanced for, while, do-while) is limited to at most $n$ executions (default $n$ = 3). This assures that there are no infinite loops, as not all loops have well-defined exit conditions identifiable through symbolic execution. This avoids repeating the behaviour of the loop when represented as a PIDG under such circumstances. Secondly, any recursive method will execute at most $m$ times (default $m$ = 2). This is to ensure that there is no potential for infinite recursion. This is often seen when calculating a value or searching for a value with a recursive algorithm. Thirdly, no object or field will ever be \textit{null}, unless \textit{null} is explicitly assigned. This is to account for cases where fields have an assigned value that is not exposed to the symbolic executor. Subsequently, when a \textit{null} field is encountered, a symbolic value is created in its place. Finally, array indexes are never out of bounds, and will always contain a value (i.e. will never be \textit{null}). This is to account for situations where an array index is accessed by a symbolic value. In such situations the return value is ambiguous, and hence a symbolic value is returned.}

\subsection{Representing Program Behaviour}

\begin{table}
    \centering
    \caption{Description of PIDG elements}

    \begin{tabular}{|p{2cm}|p{5.5cm}|}
         \hline
         Node Type & Description \\
         \hline
         \hline
         Data & \\
         - Object & An instance of a Java class. \\
         - Array & An array of \textit{Data} values. \\
         - Value & A primitive value (e.g. of type int, double). \\
         \hline
         Operator & A primary operator (arithmetic or logical). \\
         \hline
         Method call & The invocation of an API method call. \\
         \hline
         Entry point & The entry point to the program (e.g. `main'). \\
         \hline
    \end{tabular}
    
    \begin{tabular}{c}
    \\
    (a) Node types represented on PIDG. \\
    \\
    \end{tabular}


    
    \begin{tabular}{|p{2cm}|p{5.5cm}|}
        \hline
        Edge Type & Description \\
        \hline
        \hline
        Scope & The scope of a \textit{Method Call}.\\
        \hline
        Parameter & A parameter to a \textit{Method Call}.\\
        \hline
        Supplied & The return value of a \textit{Method Call}.\\
        \hline
        Aggregation & Represents the ownership of data by storing the referred \textit{Data} as a field or array member.\\
        \hline
        Transformation & The transformation of one or more values into a new value through a primary operation.\\
        \hline
    \end{tabular}
    
    \begin{tabular}{c}
    \\
    (b) Edge types between PIDG nodes.\\
    \\
    \end{tabular}
    

    
    \begin{tabular}{|p{2cm}|p{5.5cm}|}
        \hline
        Node Attribute & Description \\
        \hline
        \hline
        Runtime type & \\
        - Object & The class type of an object. \\
        - Array & The class type and component type of the array. \\
        - Value & The primitive type of the value. \\
        - Operator & The primitive result type of the operator. \\
        \hline
        Literal & The literal value of a concrete Value node. \\
        \hline
        Value & The stack value (either concrete or symbolic) of a \textit{Value} node. \\
        \hline
        String & The String value of a Java String object (that may be literal or symbolic). \\
        \hline
        Operator & The primary operator of an \textit{Operator} node. \\
        \hline
        Method signature & The qualified method signature of an \textit{Entry point} or \textit{Method call} node. \\
        \hline
    \end{tabular}
    
    \begin{tabular}{c}
    \\
    (c) Potential attributes of PIDG nodes.\\
    \\
    \end{tabular}
    
        

    \begin{tabular}{|p{2cm}|p{5.5cm}|}
        \hline
        Data Node Flag & Description \\
        \hline
        \hline
        Static & The \textit{Data} is identified to be stored in a static field. \\
        \hline
        Symbolic & The \textit{Data} has no known concrete value, and is represented by a symbol (i.e. is a symbolic input to the program). \\
        \hline
        Concrete & The \textit{Data} has a well-defined concrete value (i.e. the value is derived from evaluating a literal expression). \\
        \hline
        Synthetic & The \textit{Data} is synthesised as a result of interaction with the execution environment (e.g. the result from an API method call). \\
        \hline
        Entry point \newline parameter & The \textit{Data} is supplied as a parameter to the \textit{Entry point} (i.e. is a direct input to the program). \\
        \hline
    \end{tabular}
    
    \begin{tabular}{c}
    \\
    (d) Valid flags for PIDG \textit{Data} nodes.\\
    \end{tabular}
    

    
    \label{tab:pidg_des}
\end{table}


BPlag utilises the Program Interaction Dependency Graph \cite{cheers2020-icsim} (PIDG) to represent program behaviour for evaluating similarity. The PIDG combines the representation of: data and its relations, the transformation of data, and interactions with the execution environment into a single graph-based format. A PIDG is defined as a labelled directed graph, $G = (N, E)$, where $N$ is the set of labelled nodes and $E$ is the set of directed edges. All nodes on the graph are of one of four types, listed in table 2a. The edges between PIDG nodes represent how data is utilised in the execution of the program. These represent the dependencies between data (how it is composed), the transformation of data and interactions with the execution environment. Edges conform to one of 5 types, listed in table 2b. The nodes of the PIDG contain semantic attributes to aid in the analysis of the graph for similarity. These attributes attempt to provide context to how the node is semantically used in the program. The list of possible attributes are listed in table 2c. To enable fine-grained analysis of the graph in determining the scoping and life-cycle of data, life-cycle flags are recorded for \textit{Data} nodes. Valid \textit{Data} node flags are listed in table 2d.

By expressing program behaviour as PIDGs, linearity is removed from the representation of a program. All programs are inherently executed in sequence. However, the behaviour of a program does not need to be represented linearly. By removing linearity, it allows BPlag to analyse program behaviour irrespective of how the program is structured.

It is argued that this method of program representation is sufficient to correctly differentiate between maliciously and innocently similar programs. It is unlikely that given two students working on a complex assignment in parallel that they will implement it with the same data model (i.e relations between data). Likewise, it is unlikely they will process data in the same form, and subsequently write a program that utilises the same library calls identically. It is expected that two unrelated programs may exhibit similarity in these three aspects, however, it is not expected that any two given programs will have identical representations unless they have been plagiarised. 


\subsubsection{Graph Construction}

\begin{table*}[tbp]
    \centering
    \caption{Mapping of execution trace events to PIDG elements.}
    \label{tab:recordmapping}
    \begin{tabular}{|p{2.5cm}|p{14cm}|}
         \hline
         {Event} & {Mapping}\\
         \hline
         \hline
         API Method Call & A \textit{Method call} node $n$ is constructed for the API call, labeled by the record's method signature. A \textit{Scope} edge is created from the \textit{Scope} value to $n$ for non-static methods. \textit{Supplies} edges are created from each \textit{Parameter} to $n$. If the method returns a value, a \textit{Supplies} edge is created from $n$ to the \textit{Result} that is labeled as \textit{Synthetic}. \\
         \hline
         Field Assignment \& \newline Retrieval & For static fields, the assigned \textit{Value} is flagged as being \textit{Static} scoped. For instance fields, an \textit{Aggregation} edge is created between the \textit{Scope} and \textit{Value} indicating ownership. The edge is labeled by the \textit{Field Name}. \\
         \hline
         Array Assignment \newline \& Retrieval & An \textit{Aggregation} edge is created between the \textit{Scope} array and \textit{Value} indicating ownership. The edge is labeled by the element \textit{Index}. \\
         \hline
         Primary \newline Operation & An \textit{Operator} node $n$ is constructed, labeled by the applied \textit{Operator}. \textit{Transformation} edges are created from the \textit{LHS} \& \textit{RHS} data to $n$. A \textit{Transformation} edge is created between $n$ and the \textit{Result} data. \\
         \hline
         String \newline Concatenation & An \textit{Operator} node $n$ is constructed, labeled as \textit{Concatenation}. \textit{Transformation} edges are created from the \textit{LHS} \& \textit{RHS} data to $n$. A \textit{Transformation} edge is created between $n$ and the \textit{Result} data. \\
         \hline
         Stringification & A \textit{Transformation} edge is created between the \textit{LHS} and \textit{Result} data nodes. \\
         \hline
         Assertion & None - \textit{Assertions} \hl{are not represented in a PIDG, but may be used for program analysis.} \\
         \hline
    \end{tabular}
\end{table*}

A single PIDG is created for each execution trace returned from the symbolic executor component. Each PIDG represents the behaviour of a program over a single execution. \hl{This is in a `flattened' manner, representing the utilised data and relations, its transformation, and interactions with the execution environment. An individual PIDG does not describe the branching of a program. However, a set of constructed PIDGs represents the execution behaviour of a program over all possible executions.} 

The PIDG is constructed by mapping runtime events (recorded by the symbolic executor) to graph components. Table \ref{tab:recordmapping} describes the mapping of each event to elements in the constructed graph. The first node created in any graph is the \textit{Entry point} node. This is followed by the inclusion of any entry point parameters. The recorded runtime events are subsequently mapped to graph elements, representing program behaviour in terms of data and its relations, how it is transformed, and interactions with the execution environment. \textit{Object}, \textit{Array} and \textit{Symbolic Value} nodes appear once per graph (as they are unique). \textit{Concrete Value} nodes are duplicated on each reference, allowing for the same primitive value to appear multiple times. \textit{API method call} and \textit{Operator} nodes are created for each distinct usage.

\subsubsection{Worked Example}

\begin{figure*}[tbp]
    \begin{tabular}{p{0.4\textwidth}p{0.6\textwidth}}
        \begin{minipage}{.4\textwidth}
        \begin{lstlisting}
public static String hashPassword (
    User user, HashFunction hasher
) {
    String password = user.getPassword();
    String salt = user.getSalt();
    String input = password + salt;
    String hash = hasher.hash(input);
    return hash;
}
        \end{lstlisting}
        \end{minipage}
        &
        \begin{minipage}{.6\textwidth}
        \centering
        \includegraphics[width=0.9\linewidth]{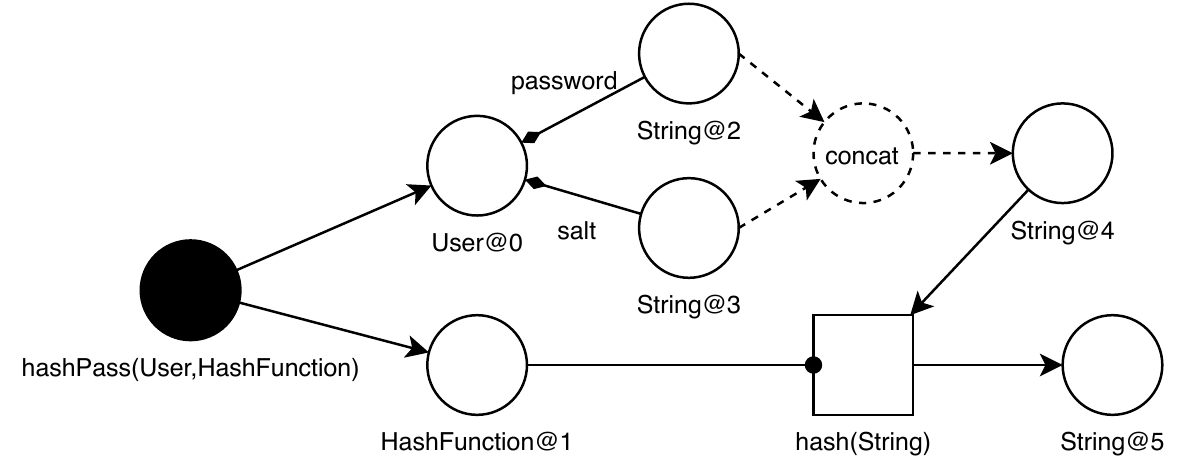}
        \end{minipage}
    \end{tabular}
    \caption{Example of a password hashing function (left) when represented as a PIDG (right). Solid node indicates the entry point. Empty nodes indicate \textit{Data}. Dashed node indicates \textit{Operator}. Square node indicates \textit{API method call}. Solid arrow indicates \textit{Supplies}. Dashed arrow indicates \textit{Transform}. Diamond arrow indicates \textit{Aggregation}. Round arrow indicates \textit{Scope}.}
    \label{fig:workedexample}
\end{figure*}

Fig. \ref{fig:workedexample} exemplifies representing a program as a PIDG. This example exemplifies the hashing of a password and with an appended salt value. A User object (\textit{User@0}) and HashFunction object (\textit{HashFunction@1}) are passed to the method call `hashPassword' as entry point parameters. The fields `password' (\textit{String@2}) and `salt' (\textit{String@3}) are implicitly retrieved from the `user' object (\textit{User@0}). Note, it is assumed the \textit{User} class is defined in source code, and hence the getter methods are followed upon invocation. The salt password and salt values are then string concatenated (i.e. the `+' operator) into the variable `input' (\textit{String@4}). This value is then passed to the `hash' function (\textit{hash(String)}), resulting in the hashed representation stored in variable `hash' (\textit{String@5}). Note, in this example all values are symbolic, and the program contains only 1 execution path as there are no branching statements. If the program contained branching statements, it would be represented by multiple PIDGs.  

\subsection{Measuring Behavioural Similarity}

\hl{
BPlag applies a bottom-up approach to evaluate the similarity of two programs. It starts by evaluating the similarity of \textit{all} PIDGs between two programs. Given programs $A$ and $B$, with PIDG sets $P_A$ and $P_B$, the pairwise similarity of PIDGs between the two programs is calculated with the PIDG similarity function, $sim_g(X, Y)$ (described in the following sub-sections). PIDGs are then mapped between programs $A$ and $B$ based on their highest evaluated similarity, i.e. PIDG $a_x \in P_A$ is mapped to PIDG $b_x \in P_B$ if $sim_g(a_x, b_x)$ is the maximum of all PIDGs contained in $P_B$. By identifying the highest similarity between two PIDGs, this is considered an indication of potentially similar execution behaviours between two programs. The mapping of PIDGs is not exclusive, meaning each PIDG may be mapped to multiple other compared PIDGs. This is to account for cases where one execution path may be executed multiple times due to plagiarism-hiding transformations.

The similarity of two programs is then calculated as the average similarity of the mapped PIDGs. This similarity value is then considered to represent the total similarity of two programs. The approach for evaluating the similarity of two PIDGs is defined in the following sub-section.
}

\subsubsection{PIDG Comparison Process}

A common method to evaluate graph similarity is to identify isomorphic sub-graphs. While accurate, sub-graph isomorphism is suffers from high computational complexity and does not scale to large graph sizes \cite{baxter1998, liu2006}. As the PIDG has the potential to contain a large number of elements, sub-graph isomorphism is not an applicable method for evaluating the similarity of PIDGs. To compare the PIDGs efficiently, an approach was developed that takes advantage of the semantics between two programs. The graph comparison approach maps data nodes between two graphs by their local usage. It starts by mapping points-of-reference between two graphs (i.e. API method calls and primary operations) and incrementally expands out, mapping nodes and edges with similar local usages between two graphs.

This approach effectively identifies common sub-graphs between two PIDGs. It is reminiscent of token-tiling (as commonly used in SCPDTs \cite{prechelt2002}). However, applies this concept to the tiling of graph nodes and edges. The similarity of two execution traces is then measured by the size of overlapping sub-graphs. This is considered to represent the quantity of behaviour in common between the programs in a single execution trace. Subsequently, the similarity of two programs can be evaluated by the overall similarity between all extracted PIDGs.

PIDGs are compared in a two-step iterative manner. Firstly, points of reference are mapped between two graphs, indicating areas that may be semantic matches. Secondly, the mapping between two graphs is iteratively expanded by exploring the graph outwards from the set of mapped nodes and subsequently expanding the mapped nodes and edges until no more nodes can be mapped. 

The usage of primary operators and API method calls are considered as points of reference between two programs. This as in any two sufficiently complex programs, it can be guaranteed there will be some usage of primary operators and API method calls. By iterating out from these points of reference, two graphs can be mapped with greater efficiency than using sub-graph isomorphism. In all comparisons, two nodes are considered potential matches if: 
\begin{itemize}
    \item They have the same node type, semantic attributes (excluding runtime type if the type is defined in source code), and flags; and
    \item They are connected by the same type of edge to the last mapped node they are connected to.
\end{itemize}

\subsubsection{PIDG Mapping Process}

Firstly, the initial mappings between the two graphs are identified. Let $A$, $B$ be the set of all operator and API call nodes in graphs $X$ and $Y$. The elements $a \in A$ and $b \in B$ are compared pairwise if they are considered a valid match. The similarity of $a$ and $b$ is then calculated by the percentage of immediate neighbours that may also constitute valid matches. This set of candidates is then mapped by the optimal assignment of their evaluated similarity scores. This results in the initial set of all mapped primary operator and API call nodes, and their immediate neighbours.

Secondly, the mappings between the two graphs are iteratively expanded. Each node $a \in M_a$ and $b \in M_b$ (where $M_a$ and $M_b$ are the sets of last mapped nodes in $X$ and $Y$ respectively) are compared pairwise for similarity. The similarity of the nodes is once again calculated by the percentage of valid mappings of immediate neighbours to the current node, that are then mapped pairwise. This process is continued at increasing depths from the initial mapped points of reference until the sets of last mapped nodes are empty (i.e. there are no more nodes that can be mapped).

The evaluated similarity of two graphs is reduced to being the intersection between the graphs. This is the number of elements that can be mapped between the graphs as a percentage of the total size of the two graphs. For example, given two graphs $X$ and $Y$ with a mapped sub-graph $X \cap Y$, the similarity of $X$ and $Y$ is evaluated as:

\[ sim_g(X,Y) = \frac{2\times |X \cap Y|}{|X|+|Y|} \]

The utilised algorithm is designed for speed as opposed to accuracy. Hence, it is a heuristic. If the initial points of reference are not mapped correctly in the first step, there is the potential that sections of the graph will not be mapped correctly. Furthermore, if there are dis-connected components of the graphs that do not contain primary operators or API calls, they will not be mapped. However, in pilot evaluations this has minimal impact on the evaluation of similarity.

\subsection{Accommodating for Plagiarism-Hiding Transformations}

It is expected that BPlag is robust against L1-L5 transformations as BPlag focuses on the behaviour of a program. This is as all such transformations \hl{are cosmetic or structural, and} do not affect the behaviour of a program. \hl{However, L6 transformations are expected to have an impact upon the approach}. L6 transformations consist of changes to decision logic (e.g. the condition applied to an if statement) and modifying expressions (e.g. changing the order of operations). BPlag is in theory, robust to L6 transformations on the condition that such transformations do not change the behaviour of the program. However, if transformations introduce functionally-equivalent but behaviourally-distinct code, BPlag is expected to see an impact on the evaluation of similarity.

Changing the decision logic of a program would not grossly affect BPlag, as long as it results in the same code executed. The only impact is if the condition value itself is modified. For example, if a relational operator is flipped \hl{as part of an if condition}, it would result in a minor difference in behaviour. By flipping relational operations, functionally equivalent code is executed, but it is not the same behaviourally as a different operator is used. However, assuming this does not affect the overall execution behaviour of a program, such transformations would have minimal impact on the overall similarity of a program. 

Conversely, modifying expressions can have a large impact on the approach. Many \hl{functionally-equivalent} transformations may be applied to change the recorded execution behaviour. For example, consider equivalent methods of reading from a file in Java using the \textit{Files}, \textit{FileReader} and \textit{FileInputStream} APIs. These are functionally equivalent but behaviourally distinct methods in terms of how they interact with the execution environment. Hence, there is potential for such transformations to evade BPlag by introducing functionally equivalent code; that subsequently has a distinct behavioural representation. \hl{However, as discussed, a vulnerability to functionally equivalent code is necessary in BPlag. If BPlag were to identify plagiarism with functionally equivalent code, it would risk identifying \textit{all} assignments implementing the same specification as being plagiarised.} However, as discussed, it is expected such transformations would have to be applied through an entire program to have a significant effect. This is considered unlikely as rewriting a program to be functionally equivalent but behaviourally distinct would be a time-intensive task for a skilled or lazy plagiariser, and would require a level of programming skill above that expected from an inept plagiariser. 

\section{Evaluation}\label{s:ev}

\subsection{Design}

The performed evaluation is to show that in the presence of pervasive plagiarism-hiding transformations BPlag is both robust and accurate; as well that it has a reasonable level of performance in comparing programs. This evaluation is broken down into three experiments. Experiment 1 evaluates the approach for robustness to pervasive plagiarism-hiding transformations. Experiment 2 evaluates the approach for accuracy in detecting suspicious program pairs transformed with pervasive plagiarism-hiding transformations. Experiment 3 evaluates the approach for efficiency in the time taken to evaluate program similarity. 

\subsubsection{Compared SCPDTs}

BPlag is compared against 5 commonly utilised SCPDTs. Ideally, the approach would be evaluated against as many SCPDTs as possible to ensure a thorough comparison. However, as identified in prior works \cite{novak2019,cheers2020}, not all SCPDTs are made available by their authors after publication. For this evaluation, only 6 SCPDTs were identified as accessible for reuse: MOSS \cite{schleimer2003}, JPlag \cite{prechelt2002}, Plaggie \cite{ahtiainen2006}, Sim \cite{grune1989}, Sherlock-Warwick \cite{joy1999}, Sherlock-Sydney \cite{pikeunknown}.

From these 6 tools, only 5 could be consistently used for the evaluation of similarity. Unfortunately, MOSS proved to be unreliable in its use, as it cannot be run locally (MOSS is only made available as a web service). In initial evaluations, large data sets would cause MOSS to perform unreliably when evaluating the pairwise similarity of all submissions in the data sets. It was found that MOSS would hang for large intervals, and subsequently fail to respond. Hence, it could not be included in this evaluation. \hl{However, in a prior study \cite{cheers2020}, MOSS performed with poorer results than JPlag, but greater results than Sim; hence similar results can be assumed in its absence.}

JPlag \cite{prechelt2002} operates by applying a token tiling algorithm to cover one source code file with tokens extracted from another. If two source files have a large degree of coverage, they can be considered similar and hence a candidate for plagiarism. First, source code files are converted into a stream of tokens. JPlag uses its own set of tokens that abstract standard language tokens to avoid matching the same token with different meanings. Second, extracted tokens are compared between files to determine similarity by the Running-Karp-Rabin Greedy-String-Tiling algorithm \cite{karp1987} where tokens from one file are covered over another within a tolerance of mismatch. Program similarity is evaluated as the percentage of tokens from one program that can be tiled over another program.

Plaggie \cite{ahtiainen2006} is a tool that is claimed to operate similarly to JPlag. However, it is an entirely local application, compared to JPlag that was originally provided as a web service. No known publication describes the operation of Plaggie; however, from examining its source code, it operates upon tokenised representations of code evaluating similarity by token tiling. 

Sim \cite{grune1989} analyses programs for structural similarity through the use of string alignment. For two programs, Sim will first parse the source code creating a parse tree. The tool will then represent the parse trees as strings and align them by inserting spaces to obtain a maximal common subsequence of their contained tokens. The similarity of programs is then evaluated as the number of matches.

Sherlock-Warwick \cite{joy1999} (Sherlock-W) implements both text and tokenised comparison methods. In the tool, a pair of programs are compared for similarity 5 times: in their original form, whitespace removed, comments removed, whitespace and comments removed, and as a tokenised source file. In all cases, the comparisons measure similarity through the identification of `runs' - a sequence of lines common to two files that may be interrupted by anomalies (e.g. extra lines).

Sherlock-Sydney \cite{pikeunknown} (Sherlock-S) analyses programs for lexical similarity. Digital signatures of source code are generated by hashing string token sequences (not lexical tokens) extracted from text files. The digital signatures are then compared, with the similarity of files being evaluated as the number of digital signatures in common.

\subsubsection{Data Set Generation}

Accessibility of test data is a known problem in SCPDT evaluations \cite{novak2019, cheers2020}. The only authentic data sets for evaluating SCPDTs are collections of undergraduate programming assignments. However, such data sets are inherently unreliable as there is no ground truth and there is no guarantee for a diverse range of plagiarism-hiding transformations to be present.

\begin{table}[]
    \centering
    \caption{SPPlagiarise Source Code Transformations}
    \label{tab:spplagiarise-trns-operations}
    \begin{tabular}{|l|l|}
        \hline
        {Level} & {Source Code Transformation} \\
        \hline
        \hline
        \multirow{1}{*}{L1} 
        & Reformat Source Code \\
        \hline
        \multirow{5}{*}{L2} 
        & Rename identifiers (packages, classes, fields, methods, locals) \\
        & Qualify/de-qualify type names \\
        & Replace static field access with static import \\
        & Replace static method call with static import \\
        & Plus all L1 transformations \\
        \hline
        \multirow{8}{*}{L3} 
        & Change access modifiers \\
        & Move field assignment to initialiser block \\
        & Declare redundant constants \\
        & Add synchronised modifier \\
        & Assign default value to variable declaration \\
        & Move variable declarations to start of block \\
        & Rearrange member (methods, fields) declarations \\
        & Plus all L2 transformations \\
        \hline
        \multirow{2}{*}{L4} 
        & Extract block to new method \\
        & Plus all L3 transformations \\
        \hline
        \multirow{7}{*}{L5} 
        & Expand combined assignment expression \\
        & Expand prefix/postfix expression \\
        & Replace for statement with while loop \\
        & Replace switch statement with if statements \\
        & Replace literal value with static constant \\
        & Surround expression with brackets \\
        & Plus all L4 transformations \\
        \hline
    \end{tabular}
\end{table}

To accommodate for these issues, BPlag will be evaluated upon generated test data sets. This will allow for a comprehensive evaluation of the presented approach against a diverse range of plagiarism-hiding transformations. Cheers, Lin and Smith \cite{cheers2019} developed the plagiarism detection data set generation tool SPPlagiarise that applies a randomly selected set of source code transformations a random number of times at randomly selected locations in a body of source code to simulate plagiarism. All implemented transformations (listed in table \ref{tab:spplagiarise-trns-operations}) are representative of undergraduate programmers, and conform to the first 5 levels of Faidhi and Robinson's \cite{faidhi1987} transformation taxonomy.

For this evaluation, the tool has been extended to afford the configuration of how pervasively source code transformations are applied to the generated data sets. This is to replace the random number of times transformations are applied, allowing a semi-fixed number of time search transformation is applied. The pervasiveness is represented as a \textit{transformation chance}, a percentage value indicating how likely a transformation will be applied at any valid location in code. This allows for the evaluation of SCPDTs with different intensities of plagiarism-hiding transformations.

\subsection{Experiment 1: Robustness to Transformation}\label{s:ev1}

The purpose of this experiment is to demonstrate that BPlag is more robust to pervasive plagiarism-hiding transformations than the five compared SCPDTs. Robustness is considered to be the ability of a SCPDT to withstand source code transformations. A more robust SCPDT will experience a lower decrease in evaluated similarity due to plagiarism-hiding transformations, compared to a less robust SCPDT. This experiment does not evaluate robustness to individual plagiarism-hiding transformations, instead, evaluates robustness on average to each level of plagiarism-hiding transformation. 

\subsubsection{Data Set}

Two groups of programs are used to evaluate robustness. Group 1 consists of 5 undergraduate assignment submissions implemented in Java that have been collected from GitHub (previously used in \cite{cheers2020}). Group 2 consists of a collection of 24 samples of 4 families of algorithms implemented in Java. These algorithm implementations consist of: 6 change-making (2 dynamic programming, 2 iterative, 2 recursive); 6 string search (2 Boyer-Moore, 2 Knuth-Morris-Pratt, 2 Rabin-Karp); 6 sorting (2 bubble sort, 2 merge sort, 2 quick sort); and 6 Minimum Spanning Tree (2 Kruskal's, 2 Prim's, 2 reverse delete).

\subsubsection{Method}
 
Firstly, test data is generated using SPPlagiarise. The 29 test programs (from data sets 1 and 2) are used as \textit{`base'} programs for the generation of simulated plagiarised \textit{`variant'} programs. Variant programs are generated with varying configuration parameters, changing the levels of applied transformations, and pervasiveness of applied transformations. For each base program, 3 sets of 150 simulated plagiarised variants are generated, consisting of: 10 L1 variants, 20 L2 variants, 30 L3 variants, 40 L4 variants, and 50 L5 variants. Each set of variants is generated with a transformation chance of: 20\%, 40\% and 60\%. This results in 450 variant programs for each base program (13,050 variants in total), enabling the measurement of SCPDT robustness to various source code transformations with varying pervasiveness. 

The number of variants generated at each transformation level is increased to offset an increasing number of potential transformations to be selected at each level. A large number of variants are generated to minimise the impact of the randomness in the data set generation method. This reduces the impact of any outliers (i.e. inconsistent decreases in similarity caused by specific transformations) and allows for the evaluation of tool robustness to transformation on average. Higher transformation chances could be applied for generated test data, however, the 3 utilised values are sufficient for this experiment. 

Secondly, robustness is evaluated by measuring the average decrease in similarity of each variant compared to its base program. The average decrease in similarity is considered to represent the vulnerability of a SCPDT to transformation. Hence, a SCPDT that shows a lesser decrease in similarity is more robust than a SCPDT with a greater decrease in similarity. An initial reading of similarity is first measured for each base program by comparing it against itself with each SCPDT. The similarity of each variant against its base program is then measured. Subsequently, the decrease in similarity of each variant from the initial similarity reading is calculated. All decreases are then summed and averaged based on the combination of transformation level and transformation chance the variant was generated with. This average decrease in similarity is then considered to represent the vulnerability of a tool to the applied plagiarism-hiding transformations. 

The initial similarity value is considered to be the highest possible similarity value each SCPDT can measure for a base program compared to its variants. However, most SCPDTs (BPlag included) utilise heuristics to afford fast comparison and hence may not measure a 100\% initial similarity. A SCPDT will not always evaluate a program as being 100\% similar to itself, hence this value offsets any errors

\subsubsection{Results}

\begin{figure*}
    \centering
    
    \begin{tabular}{@{}c@{}c@{}c@{}}
    
    \begin{tikzpicture}[font=\footnotesize]
        \begin{axis}[
            title={20\% Transformation Chance},
            height=2.2in,
            width=2.6in,
            xlabel={Transformation Level},
            ylabel={Avg. Drop in Similarity (\%)},
            ymin=0.0, ymax=100.0,
            symbolic x coords={L1,L2,L3,L4,L5},
            legend=none,
            ]
    
            \addlegendentry{BPlag}
            \addplot [dashed,color=black,mark=star,mark options={solid}] coordinates {
                (L1,0.30)
                (L2,0.37)
                (L3,2.46)
                (L4,2.77)
                (L5,2.54)
            };
            
            \addlegendentry{JPlag}
            \addplot [color=black,mark=*] coordinates {
                (L1,0.66)
                (L2,0.68)
                (L3,2.91)
                (L4,3.97)
                (L5,5.21)
            };
            
            \addlegendentry{Plaggie}
            \addplot [color=black,mark=square] coordinates {
                (L1,5.70)
                (L2,3.33)
                (L3,6.17)
                (L4,7.54)
                (L5,9.20)
            };
            
            \addlegendentry{Sim}
            \addplot [color=black,mark=triangle] coordinates {
                (L1,2.05)
                (L2,1.33)
                (L3,4.11)
                (L4,4.64)
                (L5,6.11)
            };
            
            \addlegendentry{Sherlock-W}
            \addplot [color=black,mark=diamond] coordinates {
                (L1,12.65)
                (L2,11.66)
                (L3,15.79)
                (L4,16.99)
                (L5,20.59)
            };
            
            \addlegendentry{Sherlock-S}
            \addplot [color=black,mark=o] coordinates {
                (L1,70.77)
                (L2,70.25)
                (L3,71.02)
                (L4,71.26)
                (L5,71.54)
            };
            
            \legend{};
    
        \end{axis}
    \end{tikzpicture} &
         
     \begin{tikzpicture}[font=\footnotesize]
        \begin{axis}[
            title={40\% Transformation Chance},
            height=2.2in,
            width=2.6in,
            xlabel={Transformation Level},
            ymin=0.0, ymax=100.0,
            symbolic x coords={L1,L2,L3,L4,L5},
            legend=none,
            ]
    
            \addlegendentry{BPlag}
            \addplot [dashed,color=black,mark=star,mark options={solid}] coordinates {
                (L1,0.64)
                (L2,0.62)
                (L3,8.97)
                (L4,9.71)
                (L5,9.34)
            };
            
            \addlegendentry{JPlag}
            \addplot [color=black,mark=*] coordinates {
                (L1,0.66)
                (L2,0.77)
                (L3,8.61)
                (L4,13.12)
                (L5,17.90)
            };
            
            \addlegendentry{Plaggie}
            \addplot [color=black,mark=square] coordinates {
                (L1,3.94)
                (L2,3.04)
                (L3,13.43)
                (L4,19.18)
                (L5,24.76)
            };
            
            \addlegendentry{Sim}
            \addplot [color=black,mark=triangle] coordinates {
                (L1,0.97)
                (L2,2.38)
                (L3,12.37)
                (L4,15.18)
                (L5,22.43)
            };
            
            \addlegendentry{Sherlock-W}
            \addplot [color=black,mark=diamond] coordinates {
                (L1,13.10)
                (L2,14.78)
                (L3,28.52)
                (L4,34.02)
                (L5,43.97)
            };
            
            \addlegendentry{Sherlock-S}
            \addplot [color=black,mark=o] coordinates {
                (L1,76.76)
                (L2,77.52)
                (L3,80.05)
                (L4,80.66)
                (L5,82.03)
            };
            
            \legend{};
    
        \end{axis}
    \end{tikzpicture} &
         
     \begin{tikzpicture}[font=\footnotesize]
        \begin{axis}[
            title={60\% Transformation Chance},
            height=2.2in,
            width=2.6in,
            xlabel={Transformation Level},
            ymin=0.0, ymax=100.0,
            symbolic x coords={L1,L2,L3,L4,L5},
            legend=none,
        ]
    
            \addlegendentry{BPlag}
            \addplot [dashed,color=black,mark=star,mark options={solid}] coordinates {
                (L1,0.89)
                (L2,1.01)
                (L3,18.16)
                (L4,18.96)
                (L5,18.11)
            };
            
            \addlegendentry{JPlag}
            \addplot [color=black,mark=*] coordinates {
                (L1,0.66)
                (L2,0.88)
                (L3,20.65)
                (L4,27.99)
                (L5,37.91)
            };
            
            \addlegendentry{Plaggie}
            \addplot [color=black,mark=square] coordinates {
                (L1,3.33)
                (L2,3.04)
                (L3,26.55)
                (L4,35.88)
                (L5,46.73)
            };
            
            \addlegendentry{Sim}
            \addplot [color=black,mark=triangle] coordinates {
                (L1,0.97)
                (L2,4.41)
                (L3,26.02)
                (L4,29.61)
                (L5,46.66)
            };
            
            \addlegendentry{Sherlock-W}
            \addplot [color=black,mark=diamond] coordinates {
                (L1,13.07)
                (L2,18.18)
                (L3,45.45)
                (L4,51.77)
                (L5,69.40)
            };
            
            \addlegendentry{Sherlock-S}
            \addplot [color=black,mark=o] coordinates {
                (L1,82.59)
                (L2,82.67)
                (L3,87.24)
                (L4,87.58)
                (L5,89.44)
            };
            
            \legend{};
    
        \end{axis}
    \end{tikzpicture} \\
    
    \multicolumn{3}{c}{
    \begin{tikzpicture}[font=\footnotesize]
        \begin{axis}[
            hide axis,
            xmin=0,
            xmax=0,
            ymin=0,
            ymax=0,
            legend style={draw=white!15!black,legend cell align=left},
            legend columns = 6
        ]
        
        \addlegendimage{dashed,color=black,mark=star,mark options={solid}}
        \addlegendentry{BPlag}
        
        \addlegendimage{color=black,mark=*}
        \addlegendentry{JPlag}
        
        \addlegendimage{color=black,mark=diamond}
        \addlegendentry{Sherlock-W}
        
        \addlegendimage{color=black,mark=square}
        \addlegendentry{Plaggie}
        
        \addlegendimage{color=black,mark=triangle}
        \addlegendentry{Sim}
        
        \addlegendimage{color=black,mark=o}
        \addlegendentry{Sherlock-S}
    
        \end{axis}
    \end{tikzpicture}
    } 

    \end{tabular}
    
    \caption{Average drop in similarity evaluated by each SCPDT for variants generated with each level of source code transformation. Lower values indicate greater robustness to source code transformations.}
    
    \label{fig:ex1_robustness_comparison}
\end{figure*}
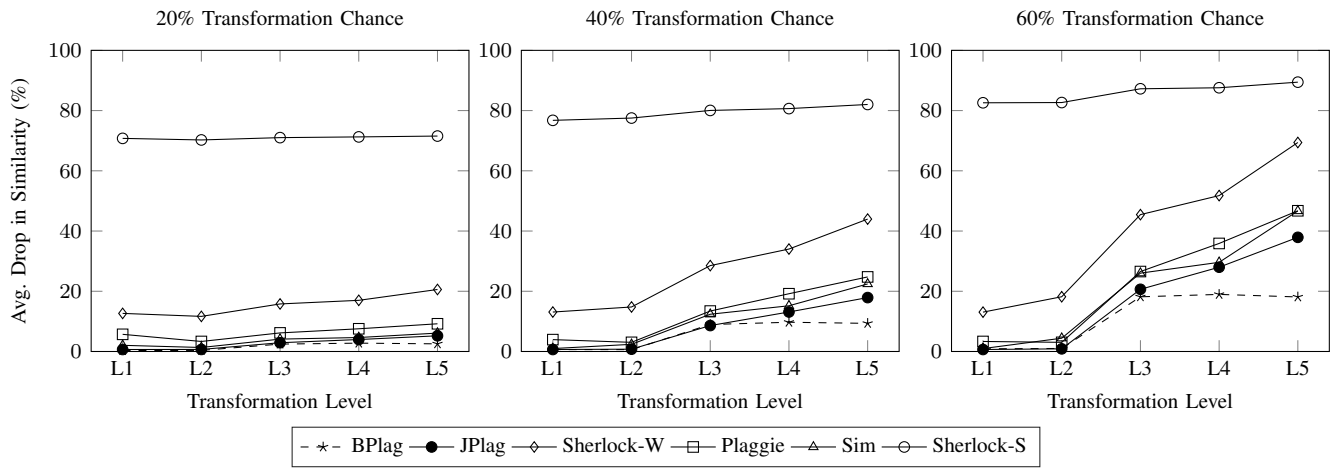

Fig. \ref{fig:ex1_robustness_comparison} compares the average drop in similarity as evaluated by each SCPDT for variants generated with the 5 levels of transformation. Lower average drops in similarity are considered to represent greater robustness to source code transformation. Over the three transformation chances, two trends can be observed. Firstly, as the chance of transformation increases, the average drop in similarity increases for each level of transformation. Secondly, all tools begin to show a sudden increase in the average drop in similarity for variants constructed with L3 transformations. For all tools but BPlag, this drop continues with the L4 and L5 transformations. 

At the 20\% transformation chance, most tools show high robustness to transformation. All but Sherlock-W and Sherlock-S fall within a tight range below 10\% across all 5 levels of transformation. It can be seen that the average drop in similarity slowly increases at higher levels of transformation. However, this is not significant, attributed to the low transformation chance. At this transformation chance, there is a negligible decrease in similarity, however, BPlag is slightly more robust than all other tools. 

At the 40\% transformation chance, the tools begin to show greater vulnerability to the plagiarism-hiding transformations. For variants created with the L1 \& L2 transformations, all tools (excluding Sherlock-S and Sherlock-W) demonstrate negligible decreases in similarity. However, starting with the application of L3 transformations, the tools suffer from a sharp and noticeable decrease in measured similarity. For the five compared SCPDTs, this decrease continues at the L4 \& L5 transformations. However, for BPlag, this decrease in similarity begins to stabilise. This implies at L3, certain transformations have a large impact on the similarity of the variants as evaluated by BPlag. However, the transformations applied to the L4 \& L5 variants do not impact upon BPlag. 

The 60\% transformation chance provides clear evidence that BPlag has greater robustness to transformation. Initially, for variants created with L1 \& L2 transformations, all tools (excluding Sherlock-S and Sherlock-W) again show negligible decreases in similarity. However, for variants created with the L3 transformations, a much larger average drop in similarity is recorded for all tools. However, the greater robustness of BPlag is demonstrated against L4 \& L5 transformations. Where the compared tools see a progressively greater decrease in average similarity, BPlag maintains a near-constant average drop in similarity. Technically, BPlag shows slightly greater robustness to L4 \& L5 transformations, however, this can largely be explained by the sample size and is not significant.

In summary of these results, BPlag is shown to be more robust to plagiarism-hiding transformations. The five compared tools show greater vulnerability to transformation at all levels of transformation, with all transformation chances. This is opposed to BPlag that consistently ranks lowest in terms of the average drop in similarity. However, BPlag shows a sharp decrease in robustness to L3 transformations. The lack of decrease in robustness to L4 \& L5 transformations implies that BPlag is robust to such transformations. This is expected, as  L4 \& L5 transformations do not modify the behaviour of a program. However, from the results, BPlag is vulnerable to specific L3 transformations. On further analysis, this is a result of \textit{`value-injecting transformations'}: transformations that result in new values represented on PIDGs. Hence, a supplementary experiment is performed to investigate their impact.

\subsubsection{Robustness Without Value-Injecting Transformations}

The drop in similarity against L3 transformations is largely caused by two specific source transformations: `declare redundant constants', and `assign default value to variable declaration'. Both of these transformations result in the addition of new values represented in PIDGs. As a result of this, it introduces new data nodes to the graphs that cannot be matched to the base program. Hence, it results in a decrease in similarity. To demonstrate the impact of value-injecting transformations, the results of BPlag are analysed in further depth. This is namely to identify the average drop in similarity experienced by BPlag when analysing variants created \textit{with} value-injection transformations, and those \textit{without} value-injecting transformations.

\begin{table}
    \centering
    \caption{Average drop in measured similarity by BPlag for variants created with and without value-injecting transformations.}
    \label{tab:ex1_variants_without_injection_comp}
    \begin{tabular}{|r|c|c|c|c|c|}
        \hline
        \multirow{2}{*}{{Trans. Chance}} & \multicolumn{5}{c|}{{Avg. Drop in Similarity (\%)}} \\
        & {L1} & {L2} & {L3} & {L4} & {L5} \\
        \hline
        \hline
        \multicolumn{6}{|c|}{With Value-Injecting Transformations} \\
        \hline
        20\% & 0.30 & 0.37 & 2.46 & 2.77 & 2.54 \\
        40\% & 0.64 & 0.62 & 8.97 & 9.71 & 9.34 \\
        60\% & 0.89 & 1.01 & 18.16 & 18.96 & 18.11 \\
        \hline
        \hline
        \multicolumn{6}{|c|}{Without Value-Injecting Transformations} \\
        \hline
        20\% & 0.30 & 0.37 & 0.35 & 0.66 & 0.93 \\
        40\% & 0.64 & 0.62 & 0.76 & 1.75 & 2.80 \\
        60\% & 0.89 & 1.01 & 1.13 & 2.56 & 4.86 \\
        \hline
    \end{tabular}
\end{table}

Table \ref{tab:ex1_variants_without_injection_comp} compares the average drop in similarity of variants created with and without value-injecting transformations. It can be seen that there is a much larger average drop in similarity for variants created with value-injecting transformations, compared to those that do not contain value-injecting transformations. In all cases, for L3+ variants there is a substantial reduction in the drop of similarity for variants created without value-injecting transformations. Hence, it can be concluded that BPlag is much more robust to transformation compared to the SCPDTs; however, it does have a vulnerability to value-injecting transformations.

It should be noted that value injection is a vulnerability of all SCPDTs, and not just BPlag. Value-injection is analogous to the addition of source code to a plagiarised program. When a conventional SCPDT attempts to evaluate similarity in such a case, it cannot match the injected source code and hence results in a drop in similarity. In BPlag, value injection changes the representation used to compare programs for similarity (i.e. PIDGs). And hence, it results in a reduction of similarity. 

\subsubsection{Discussion of Results}

From the results in Fig. \ref{fig:ex1_robustness_comparison}, BPlag demonstrates greater robustness to plagiarism-hiding transformations. As the chance of transformation increases, BPlag demonstrates progressively greater robustness to transformation, indicated by the lower average drop in similarity. However, as identified, BPlag is vulnerable to value-injecting transformations. Without value-injecting transformations, BPlag demonstrates much greater robustness to transformation.

The simplest method of accommodating for value-injecting transformations is to simply ignore the addition of these values. They could easily be identified in the PIDG as nodes with degree 0. This effectively means the data has no relation to any other data in the program and is otherwise unused. However, this does not consider if any arbitrary node with degree 0 is placed intentionally or maliciously. Therefore, any such improvement requires further investigation.

\subsection{Experiment 2: Detection Accuracy}\label{s:ev2}

The purpose of this experiment is to demonstrate that BPlag is more accurate in the detection of pervasively transformed plagiarised assignment submissions. This will be in comparison to the accuracy of the five compared SCPDTs. The tools will be evaluated by injecting simulated plagiarised variant programs into existing data sets of undergraduate assignment submissions. The accuracy of the tools will be measured by the ability of each tool to correctly differentiate between plagiarised (i.e. the generated variants) and innocent program pairs. 

\subsubsection{Measuring Accuracy}\label{s:ev2-ident}

Accuracy is a difficult aspect of a SCPDT to measure. Contrary to as the name suggests, a SCPDT does not detect plagiarism. Instead, a SCPDT identifies indications of plagiarism \cite{joy1999}. If two programs have a high similarity, it can be considered an indication that plagiarism may be present. However, there is no consensus as to what is considered a high similarity. Hence, this has resulted in different interpretations of what is considered an indication of plagiarism for use in SCPDT evaluations, and how to evaluate SCPDTs in general \cite{novak2019}.

\begin{figure}
    \centering
    \includegraphics[scale=1.2]{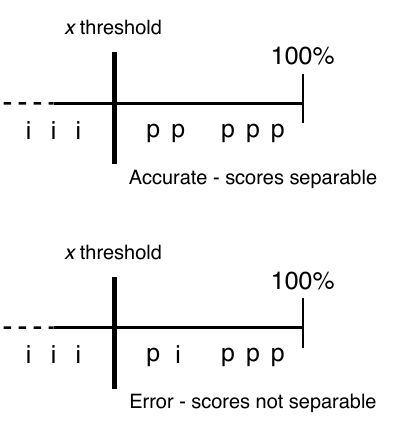}
    \caption{Example error case. If scores cannot be separated at threshold $x$ into homogeneous groups of plagiarised (p) and innocent (i), errors have occurred.}
    \label{fig:error_example}
\end{figure}

In this evaluation, the accuracy of SCPDTs is compared by the number of errors made when analysing a set of assignments. A SCPDT makes an error when it evaluates a plagiarised pair of programs as having a lower similarity score than an innocent pair of programs. Subsequently, if all similarity scores for a data set are plotted on a number line, this results in it being impossible to linearly separate those that are plagiarised and innocent into homogeneous groups. An example of this is demonstrated in Fig. \ref{fig:error_example}. On top, innocent scores (labelled i) can be cleanly separated from the plagiarised scores (p). However, on bottom, the innocent scores cannot be cleanly separated from the plagiarised scores as an innocent score is more similar than a plagiarised score. Hence, an error has occurred by measuring an innocent pair with a higher similarity than a plagiarised pair.

Using this evaluation method allows for measuring accuracy by the relative value of plagiarised and innocent similarity scores. This affords for comparing SCPDTs in terms of the number of errors they make when evaluating program similarity. A SCPDT that commits many errors is considered to be less accurate than a tool that commits few errors. Hence, the purpose of this experiment will be to show that BPlag is more accurate than other SCPDTs by committing fewer errors when detecting plagiarism.

\subsubsection{Data Set}

A collection of 12 sets of undergraduate assignment submissions are used to evaluate SCPDT accuracy. Each set of submissions are for an individual programming assignment for first and second-year programming courses. All assignment sets differ in code size, complexity, and the number of submissions. Table \ref{tab:ex3_datasets} provides an overview of this data set.

\begin{table}
    \centering
    \caption{Size of accuracy evaluation data sets.}
    \label{tab:ex3_datasets}
    \begin{tabular}{|c||c|c|c|c|}
        \hline
        {Collection} & {Submission Count} & {Avg. File Count} & {Avg. LOC} \\
        \hline
        \hline
        1 & 170 & 3.99 & 858.22 \\
        2 & 125 & 4.08 & 994.75 \\
        3 & 37 & 5.84 & 614.30 \\
        4 & 39 & 5.72 & 706.05 \\
        5 & 32 & 8.75 & 710.28 \\
        6 & 38 & 6.32 & 649.24 \\
        7 & 30 & 3.37 & 377.67 \\
        8 & 48 & 5.25 & 598.69 \\
        9 & 62 & 3.90 & 470.23 \\
        10 & 22 & 4.50 & 436.95 \\
        11 & 40 & 6.18 & 571.05 \\
        12 & 48 & 4.65 & 474.69 \\
        \hline
    \end{tabular}
\end{table}

\subsubsection{Experimental Method}

Firstly, test data is generated with SPPlagiarise. Using generated test data allows for the construction of a data set with known plagiarised programs pairs that can be injected into an existing data set of innocent assignment submissions. Hence, this affords the accurate evaluation of SCPDT accuracy with a ground-truth data set with known types of plagiarism-hiding transformations applied. This experiment will be performed with different combinations of transformations and pervasiveness of transformation. Simulated plagiarised variants will be generated with the 5 levels of transformation, as well as with 4 increasing levels of transformation pervasiveness: 40\%, 60\%, 80\% and 100\%. Higher transformation chances are used in this experiment, as experiment 1 demonstrated that BPlag is largely robust to the 20\% and 40\% transformation chances. From each of the assignment collections, 5 assignment submissions are randomly selected as base programs for the generation of simulated plagiarised variant programs. For each base program: 10 L1, 10 L2, 10 L3, 10 L4 and 10 L5 variants are created for each of the 4 chances of transformation (40\%, 60\%, 80\%, 100\%). This results in the generation of 200 variants for each selected base program, totalling 12,000 variants across the 12 data sets. This will enable a comprehensive and large scale evaluation of SCPDT accuracy.

Secondly, the SCPDTs are used to evaluate the pairwise similarity scores of all programs in the base data sets, as well as the similarity of the variants to their respective base programs. This results in 12 sets of pairwise similarity scores (one for each collection of assignments), as well as 12,000 similarity scores to be injected into these data sets. Amongst the 12 collections of assignments, there are known cases of academic misconduct present. To maintain the integrity of the experiment (in that only generated variants are considered plagiarised), all scores representing the known cases of academic misconduct are removed from the data sets. This was through manual review of the similarity scores reported by each SCPDT.

Thirdly, the error counts of the SCPDTs are evaluated. The scores of the simulated plagiarised variants (for each transformation level and chance combination) are placed into their respective base data sets. The scores are then sorted, and the error count is evaluated. The error count is evaluated by identifying the best possible separation between the plagiarised and innocent scores (i.e. when the lowest number of errors occur). This results in a worst-case of 600 errors for each transformation level and transformation chance combination (i.e. the number of variants generated). Subsequently, any innocent pairs above and the score separation point, and any plagiarised scores below this point are considered errors. 

\subsubsection{Results}

\begin{table}
    \centering
    \caption{Error counts of each SCPDT detecting simulated plagiarism. Lower counts indicate higher accuracy.}
    \label{tab:ex2_results}
    \begin{tabular}{|r||c|c|c|c|c||c|c|}
        \hline
        \multirow{2}{*}{{Tool}} & \multicolumn{5}{|c||}{{Error Count at Trans. Level}} & \multicolumn{2}{c|}{Total Errors} \\
        & {L1} & {L2} & {L3} & {L4} & {L5} & Count & Perc (\%) \\
        \hline
        \hline
        
        \multicolumn{8}{|c|}{40\% Transformation Chance} \\
        \hline
        BPlag       & 0     & 0 & 4 & 1 & 12 & 17 & \textbf{0.57} \\
        JPlag       & 31    & 34 & 51 & 56 & 27 & 199 & 6.63 \\
        Plaggie     & 111   & 111 & 198 & 226 & 222 & 868 & 28.93 \\
        Sim         & 0     & 0 & 2 & 3 & 41 & 46 & 1.53 \\
        Sherlock-W  & 0     & 2 & 19 & 18 & 57 & 96 & 3.2 \\
        Sherlock-S  & 498   & 503 & 505 & 507 & 518 & 2,531 & 84.37 \\
        \hline\hline
        
        \multicolumn{8}{|c|}{60\% Transformation Chance} \\
        \hline
        BPlag &	0 & 0 & 12 & 14 & 28 & 87 & \textbf{1.80} \\
        JPlag &	31 & 38 & 22 & 50 & 94 & 235 & 7.83 \\
        Plaggie &	111 & 111 & 291 & 327 & 360 & 1200 & 40.00 \\
        Sim &	0 & 0 & 22 & 34 & 206 & 262 & 8.73 \\
        Sherlock-W &	0 & 12 & 121 & 165 & 341 & 639 & 21.30 \\
        Sherlock-S &	522 & 529 & 535 & 539 & 555 & 2,680 & 89.33 \\
        \hline\hline
        
        \multicolumn{8}{|c|}{80\% Transformation Chance} \\
        \hline
        BPlag &	0 & 0 & 50 & 48 & 60 & 158 & \textbf{5.23} \\
        JPlag &	31 & 38 & 92 & 161 & 279 & 601 & 20.03 \\
        Plaggie &	111 & 111 & 413 & 468 & 507 & 1,610 & 53.67 \\
        Sim &	0 & 2 & 85 & 119 & 441 & 647 & 21.57 \\
        Sherlock-W &	0 & 35 & 331 & 389 & 534 & 1,289 & 42.97 \\
        Sherlock-S &	553 & 554 & 576 & 575 & 578 & 2,836 & 94.53 \\
        \hline\hline
        
        \multicolumn{8}{|c|}{100\% Transformation Chance} \\
        \hline
        BPlag &	0 & 0 & 88 & 90 & 110 & 288 & \textbf{9.60} \\
        JPlag &	31 & 41 & 214 & 304 & 483 & 1,073 & 35.77 \\
        Plaggie &	111 & 111 & 525 & 573 & 571 & 1,891 & 63.03 \\
        Sim &	0 & 24 & 160 & 197 & 575 & 956 & 31.87 \\
        Sherlock-W &	1 & 70 & 521 & 540 & 582 & 1,714 & 57.13 \\
        Sherlock-S &	582 & 582 & 582 & 582 & 582 & 2,910 & 97.00 \\
        \hline
    \end{tabular}
\end{table}

Table \ref{tab:ex2_results} lists the error counts for each SCPDT for detecting the simulated plagiarism. This is broken down into error counts at each level of transformation and chance of transformation combination, and are aggregated for all data sets for ease of review. From the results, BPlag is much more accurate in the presence of pervasive plagiarism-hiding transformations. For example, at the 40\% transformation chance, BPlag has cumulatively 17 errors over the 5 levels of transformation. This is compared to the next closest tool, Sim, that has cumulatively 46 errors recorded. At the 60\%, 80\% and 100\% transformation chances, the error counts of all tools, including BPlag, continue to increase. However, BPlag experiences the least increase in errors out of all the tools. Hence, it commits the fewest errors when detecting the simulated plagiarised variants.

All errors in this experiment result from submission pairs being incorrectly identified as plagiarised (for programs originally contained in the 12 data sets) or innocent (for the generated variants). This is caused by the similarity scores of some variant programs dropping to a point where correctly innocent scores are ranked higher. This effectively results in the similarity scores of the plagiarised and innocent program pairs becoming no longer linearly separable. Hence, there are unavoidable errors that occur when identifying the plagiarised programs. Note, the L3+ variants in this experiment were generated with value-injecting transformations. Hence, it was expected errors would occur as this is a known vulnerability of BPlag. 



\subsubsection{Discussion of Results}

From the results in table \ref{tab:ex2_results}, BPlag is shown to be the most accurate in detecting the simulated cases of plagiarism across the six compared SCPDTs. By analysing program behaviour, it is capable of consistently measuring a higher similarity between the simulated plagiarised programs, as compared to the five compared SCPDTs. Overall, this has resulted in a lower error count at all level of transformation and transformation chance combinations. While BPlag does suffer a decrease in accuracy when detecting the L3+ variants (in theory BPlag should be not affected by such transformations), this is explained by the value-injecting transformations identified in experiment 1. For L3+ transformations applied with higher transformation chances (i.e. are more pervasively applied), this appears to have a profound impact on accuracy. However, the results of BPlag at the 100\% transformation chance, are more similar to the five compared tools results at the 40\% or 60\% transformation chances. Hence, even in the presence of such transformations, BPlag remains the most accurate tool in this experiment.


The results roughly correspond to the robustness evaluation in Fig. \ref{fig:ex1_robustness_comparison}. The SCPDTs will generally see a small decrease in similarity against L1 and L2 transformations. This is reflected in this experiment by a low error rate against L1 and L2 transformations (that are largely consistent across the 4 transformation levels, excluding Sherlock-S and Sherlock-W). Furthermore, the compared tools started seeing progressively larger decreases of accuracy against the L3+ transformations. This correlates with the large average drop in similarity when exposed to L3+ transformations in Fig. \ref{fig:ex1_robustness_comparison}. Hence, these results are largely consistent with experiment 1. Comparatively, BPlag is subject to a progressively smaller decrease in accuracy when exposed to L3+ transformations. 

The accuracy of the compared SCPDTs can be also be compared in general from the results. If the tools are ranked by lowest cumulative errors, BPlag is consistently first. This is followed by Sim, JPlag, and Sherlock-W that shuffle between second, third and fourth place. Plaggie and Sherlock-S are consistently ranked fifth and sixth, both having significantly higher error counts than the other tools.




\subsection{Experiment 3: Efficiency of Comparison}

This experiment will evaluate BPlag in its efficiency in measuring source code similarity. Efficiency is an important aspect of a SCPDT. It is unreasonable for a tool to be so computationally complex that it requires large periods of time to analyse a set of assignment submissions. Therefore a SCPDT must be able to evaluate a data set in a reasonable amount of time. BPlag combines two computationally intensive areas: symbolic execution and graph similarity. Therefore, it is not expected that this approach is more efficient than the five compared SCPDTs. However, this evaluation is to show that BPlag is capable of measuring similarity without an excessive run time.

This experiment was conducted on a single workstation with a 32-core/64-thread CPU with 128GB of RAM running Ubuntu 20.04 LTS. All tools are executed in parallel. For all Java-based tools, Java version 11 is utilised with the GraalVM\footnote{https://www.graalvm.org} Java Virtual Machine. GraalVM is used as opposed to the standard Oracle HotSpot Java Virtual Machine as the authors have observed that it has a lower peak memory usage in long-running experiments, or when analysing large data sets.

\subsubsection{Method}

This experiment is controlled by two scripts written in Kotlin (a Java-like programming language). Both scripts handle the invocation of the SCPDTs on a pair of programs and monitor the time each SCPDT spends executing on each data set. This experiment utilises the data set of experiment 2 (i.e. the collection of 12 sets of undergraduate assignment submissions) while evaluating all pairwise similarity scores of the assignment collections.

The first script manages the execution of the five compared tools. Each of these tools is exposed through a Java binding (i.e. a Java class exposing the tool) that affords procedural invocation of the tool from Java code. The script enforces that at any one time: only one data set can be analysed by a single tool, and each tool can be executed at most 32 times in parallel. Subsequently, this script records the runtime of each tool invocation and calculates the average execution time per program pair.

The second script manages the invocation of and monitors the execution time of BPlag. BPlag is not implemented as a single `tool', but as three separate programs as part of a pipeline (i.e. execution trace generation, PIDG construction and PIDG comparison, as per Fig. \ref{fig:bplag_overview}). In this setup, the outputs of the first stage (i.e. execution traces) and second stage (i.e. PIDGs) are serialised and stored on disk before being passed to the next stage. This script manages the execution of each stage and the intermediate storing of data. Like the compared tools, BPlag is executed in parallel, however, execution traces are extracted at most 32 times in parallel, PIDGs constructed at most 8 times in parallel, and PIDGs compared at most 8 times in parallel.

Many SCPDTs (e.g. JPlag) support the evaluation of all submissions in a data set simultaneously (i.e. comparing all submissions in a single execution of a tool). While other tools (e.g. Sherlock-S), only support evaluating a single pair of submissions. To ensure a consistent comparison, this experiment evaluates the efficiency of the SCPDTs in evaluating similarity on a per-submission-pair basis, i.e. the SCPDTs are executed once per submission pair in each data set. The total execution time for each tool is then summed, normalised and averaged to identify the average execution time per program pair. As multi-processing is utilised, it can distort the perceived efficiency of the tools. Hence, the execution times are subsequently normalised by the maximum number of parallel invocations. Therefore, the time taken by the compared tools is multiplied by 32, and the time taken by BPlag is multiplied by 32, 8, and 8 for its three stages (respectively); with both being averaged on a per-submission basis. This results in the normalised average execution times for each tool to evaluate the similarity of a single pair of programs. 

The construction and comparison of PIDGs can consume large amounts of memory, hence the parallelism of the last 2 phases of BPlag is limited for stability and reliability. If the three stages are amalgamated into a single program, it can result in poor efficiency. This is caused by the JVM garbage collector constantly churning to free up heap space (as the current implementation used large quantities of memory). By implementing the three stages as separate programs, it effectively allows the approach to run without any garbage collection overhead. Hence, it avoids this garbage collection churning. However, this is at the added cost of serialising and storing data between the three stages on disk. The five compared SCPDTs do not incur a garbage collection overhead, hence this does not provide an unfair advantage for BPlag.

\subsubsection{Results}

\begin{table*}
    \centering
    \caption{Average execution time of each SCPDT per submission pair (in seconds) for each collection of assignment submissions.}
    \label{tab:ex3_avg_exec_time}
    \begin{tabular}{|c|c||cccccc|}
        \hline
        \multirow{2}{*}{Collection} & \multirow{2}{*}{No. Pairs} & \multicolumn{6}{c|}{Average Execution Time per Submission Pair (seconds)} \\
         & & {BPlag} & {JPlag} & {Plaggie} & {Sim} & {Sherlock-W} & {Sherlock-S} \\
        \hline
        \hline
1 & 14,365 & 6.00 & 2.51 & 0.91 & 0.35 & 4.05 & 0.01 \\
2 & 7,750 & 6.60 & 2.57 & 0.90 & 0.20 & 4.19 & 0.24 \\
3 & 666 & 3.41 & 2.38 & 0.58 & 0.22 & 3.32 & 0.07 \\
4 & 741 & 4.43 & 2.46 & 0.71 & 0.71 & 3.76 & 0.13 \\
5 & 496 & 2.95 & 2.42 & 0.68 & 0.29 & 3.77 & 0.10 \\
6 & 703 & 3.55 & 2.39 & 0.41 & 0.89 & 3.41 & 0.07 \\
7 & 435 & 7.52 & 2.32 & 0.55 & 0.33 & 2.43 & 0.11 \\
8 & 1,128 & 6.52 & 2.43 & 0.68 & 0.30 & 3.32 & 0.21 \\
9 & 1,891 & 4.97 & 2.36 & 0.66 & 0.76 & 2.74 & 0.05 \\
10 & 231 & 3.98 & 2.29 & 0.62 & 0.62 & 2.70 & 0.21 \\
11 & 780 & 2.35 & 2.40 & 0.68 & 1.05 & 3.20 & 0.12 \\
12 & 1,128 & 2.30 & 2.38 & 0.64 & 1.02 & 2.77 & 0.13 \\
\hline
Average & - & 4.55 & 2.41 & 0.67 & 0.56 & 3.30 & 0.12 \\
\hline
    \end{tabular}
\end{table*}

Table \ref{tab:ex3_avg_exec_time} presents the average execution times per program pair in each of the 12 collections of assignment submissions, normalised for concurrent execution. It can be seen that Sherlock-S is by far the most efficient tool on average. This is followed by Sim, Plaggie, JPlag, Sherlock-W, and finally, BPlag. BPlag is approximately 38 times slower than Sherlock-S and almost 2 times slower than JPlag. This is expected, as BPlag is considerably more computationally complex.

\subsubsection{Discussion of Results}

From these results, BPlag is not the most efficient SCPDT. The poorer efficiency is attributed to the comparatively more complex implementation of the approach. Being largely graph-based, it is expected that BPlag will require greater run time. The impact of greater complexity is minimised with the presented graph-matching heuristic that is designed to operate in approximately $O(n^2)$ time. This is the same complexity of the algorithms used by JPlag and Plaggie. However, BPlag still has a much greater run time due to the sheer number of graphs extracted to represent a single program that all require pairwise comparison. Hence, BPlag is inherently more complex, requiring a greater run time.

The greater runtime does not mean BPlag is impractical to use. If only a pair of programs are analysed, the runtime would not cause a human reviewer to wait an unreasonable amount of time. If BPlag was running on the largest data set (collection 1, 170 programs), it would take approximately 24 hrs to run single-threaded. While this is, of course, a long time for the analysis of a single (large) data set, modern CPUs are multi-cored, and as such BPlag takes advantage of this. On a modern 8-core machine, BPlag would take approximately 3 hrs to execute on collection 1. Furthermore, if a more reasonably sized data set was analysed, (e.g. 100 submissions) it would take approximately 45 mins on an 8-core machine. 

It must also be considered that there is greater benefit in a SCPDT being more accurate than efficient. A SCPDT is typically used once per assessment task to identify indications of plagiarism. In this single execution, a SCPDT needs to accurately identify any indications of plagiarism, especially in the presence of plagiarism-hiding transformations. This is afforded by BPlag. 

\section{Discussion}


The approach to SCPD utilised by BPlag was designed to be both robust to plagiarism-hiding transformations, and accurate in the detection pervasively transformed works. In doing so the approach analyses the behaviour of assignment submissions, as the behaviour of a program typically does not change due to plagiarism-hiding transformations. Greater robustness to plagiarism-hiding transformations is afforded by the approach analysing the behaviour of a program. As demonstrated in experiment 1, by analysing behaviour, the plagiarism-hiding modifications that transform the structure of source code are rendered largely inert. This is as the plagiarism-hiding transformations do not have a significant impact on the behavioural representation of a program. However, the approach does show vulnerability to transformations that modify the behaviour of a program, i.e. transforming to functionally-equivalent code or the identified value-injecting transformations. Subsequently, the approach is more accurate in the detection of plagiarism, where the plagiarised work is pervasively transformed with plagiarism-hiding transformations, as demonstrated in experiment 2. Greater accuracy is largely a result of the greater robustness, allowing for BPlag to evaluate a higher similarity between plagiarised works (with pervasive plagiarism-hiding transformations) than innocent works. However the greater accuracy is also influenced by the approach representing the unique implemented behaviour of a program, as expressed through source code. As a trade-off for greater robustness and accuracy, BPlag is less efficient than the five compared SCPDTs due to the greater complexity of the approach. However, the greater computational complexity does not make the approach infeasible to use.

\subsection{Limitations} 

The conducted evaluations rely on generated test data. Test data generation was used to create a comprehensive evaluation data set that is representative of undergraduate assignment submissions, that also can be guaranteed to contain plagiarism-hiding transformations. If the evaluation were to utilise real data sets containing plagiarism, neither of these factors could be guaranteed. i.e. in any data set of undergraduate assignment submissions, there is no guarantee that plagiarism is, in fact, present, nor that it will contain a diverse range of plagiarism-hiding transformations.

The generated test data was created with SPPlagiarise. SPPlagiarise implements a total of 19 semantics-preserving source code transformations conforming to L1-L5 of Faidhi and Robinson's \cite{faidhi1987} taxonomy. Semantics-preserving transformations were applied as BPlag requires semantically correct programs to analyse. It cannot analyse programs that are non-compilable. Hence, the evaluation required test data generation that could be guaranteed not to break a program. A limitation of this is that there are many more plagiarism-hiding transformations than the 19 implemented transformations. Hence, the results of the evaluation only show that BPlag is robust and accurate in the presence of these 19 transformations. Furthermore, SPPlagiarise does not implement L6 transformations. L6 transformations are in particular difficult to automate. It requires transforming expressions into functional equivalents (e.g. using functionally-equivalent APIs) and potentially complex rewriting of conditional expressions. Hence, extending SPPlagiarise to support L6 transformations in a meaningful manner was omitted from this work. However, it can be expected that BPlag, along with the five compared SCPDTs, will suffer a decrease in robustness and accuracy in the presence of such L6 transformations. 

As identified, BPlag is vulnerable to transformations that change the behavioural representation of a program. Such transformations are summarised as either: the introduction of functionally-equivalent code, or value-injecting transformations. As discussed, the vulnerability to functionally equivalent code is a required deficiency of BPlag to avoid errors when detecting plagiarism. However, as a generalisation, all SCPDTs are vulnerable to transformations that change the representation of a program used by the SCPDT. When the representation of a program is modified, it is inevitable any source code similarity tool will measure a decrease in similarity. From this reasoning, there is the potential for the structure-based tools to be more robust and accurate than BPlag against behaviour-modifying transformations. For example, consider the use of functionally-equivalent APIs. Code fragments implementing functionally equivalent code can be structurally similar, even though they have different semantics and behaviour. Hence, in such a case it is feasible that a SCPDTs such a JPlag would measure a higher similarity than BPlag in such cases.

\hl{There are also important limitations of the evaluation caused by the composition of the base programs used to generate test data. All data sets have very different compositions in terms of source code (i.e. the number and type of declarations, statements and expressions); and each source code transformation applied by SPPlagiarise can only be used \textit{iff} there is a valid location for it to applied in code. Hence, on data sets generated with different base programs, there is a potential for very different results to be obtained as different counts of transformations can be applied. Hence, the results can only conclusively determine the performance of BPlag on the utilised data set. However, this problem is common to all evaluations that measure source code similarity, and is mitigated in this experiment with a large and diverse data set.}

\section{Related Work}

BPlag is a distinctly behavioural approach to academic SCPD. The approach does not share similarity with existing approaches to SCPD. All known proposed SCPDTs implement structural or semantic measures of similarity. Furthermore, all known \textit{available} SCPDTs implement structural measures of similarity only. The lack of similar approaches to SCPD is supported by the recent literature reviews \cite{martins2014,novak2019} that identify various structural and/or semantic SCPDTs, however, fail to identify any behavioural SCPDTs.

The implemented approach to academic SCPD shares more in common with program plagiarism detection tools than conventional SCPDTs. Program plagiarism detection tools typically analyse program binaries or execution traces to identify `fingerprints' of similarity that may indicate a program (or a component of it) has been stolen. Such tools implement similar methods in how program similarity is analysed, for example similar tools include: Cop \cite{luo2014, luo2017}, LoPD \cite{zhang2014, ming2016}, and VaPD \cite{jhi2011, jhi2015}.

Cop identifies similar components between two programs by analysing semantics (behaviour) through symbolic execution. It is concerned with identifying basic blocks of code (i.e. a continuous sequence of instructions that do not contain branching instructions) that have the same input/output relations. Subsequently, common subsequences of semantically equivalent basic blocks can be identified between two programs, enabling the evaluation of similarity. 

LoPD identifies the similarity of programs based on their implemented program logic to determine if they are semantically equivalent. The approach utilises symbolic execution to identify the usage of data in an execution path of a program and any mathematical constraints that are placed upon it. Instead of identifying if two programs are the same under a set of inputs, the approach aims to prove that two programs are not semantically equivalent by identifying an input where the programs behave differently. If one such case can be found, it can be proven that the programs are not semantically equivalent. 

VaPD identifies program similarity through runtime execution analysis of the values stored in memory during program execution. The foundation of this approach is the observation that certain runtime values of a program cannot be changed through semantics-preserving obfuscations. These values are identified, extracted and refined to characterise a program. The values are then compared to identify overlaps and hence program similarity.

BPlag is similar to these works by measuring similarity with behavioural means. Furthermore, they utilises some similar techniques (e.g. symbolic execution) to analyse behaviour. While these works share individual aspects of similarity with BPlag, there are noticeable differences. The major difference between BPlag and these tools is the form a program takes when analysed. BPlag analyses source code to derive a behavioural representation, while the compared works analyse machine code or bytecode. There are also fundamental differences in there operations. CoP and LoPD look for elements of functional equivalence between two programs through input-output relations. This is more akin to a `black-box' method of identifying if two programs are similar. While BPlag seeks to identify similar programs by analysing how the program behaves at runtime by identifying data and its relations, transformation of data, and method calls. This is more akin to a `white-box' method of identifying if two programs are similar, explicitly looking at the logical implementation of a program. Similarly, BPlag does not seek to identify plagiarism by identifying functionally equivalent code, due to the expectation that academic assignments are expected to be largely functionally equivalent. Instead, BPlag looks for the same execution behaviour. VaPD implements a value-based fingerprinting method to characterise programs. This is a much lower-level approach compared to BPlag, and fingerprinting is different from the approach of BPlag. In comparison, fingerprinting identifies indications of plagiarism as a `subset' of a program. This is in contrast to BPlag that identifies the `global' similarity of two programs by comparing all represented execution behaviour. 

Overall, BPlag is designed as a SCPDT. Any similarity with program plagiarism detection tools is a result of analysing program behaviour. However, as seen from the results of the evaluation, utilising similar approaches has resulted in a SCPDT that is both more robust and accurate than currently available SCPDTs.

\section{Conclusion and Future Work}

In this article, the design and evaluation of BPlag, a novel behavioural approach to SCPD, has been presented. BPlag is intended to identify the most extreme cases of academic source code plagiarism, where a skilled plagiariser has pervasively transformed source code, and submitted it as their own work. The presented evaluations support that in the presence of pervasive plagiarism-hiding transformations: BPlag is both more robust to source code transformations, and more accurate in detecting the simulated plagiarised programs. However, the evaluations have also shown that BPlag is comparatively less efficient. This was expected due to the complexity of BPlag, and the potential for further optimisation (e.g. comparison pruning). 

There are numerous directions of future work for improving, extending or re-applying BPlag. Firstly, BPlag is vulnerable to value-injecting transformations. An area of future work is investigating an appropriate means for mitigating the impact of such transformations. \hl{Furthermore, BPlag needs to be evaluated against more diverse transformations, as well as data sets of real plagiarism to identify any more vulnerabilities that may occur with real data.} Secondly, different approaches to measuring similarity (i.e. structural, semantic and behavioural) provide valuable insights into how a program is similar. It would be interesting to explore if BPlag could be combined with structural and semantic tools to provide a much more comprehensive view of how two programs may be similar. This could be applied to investigate if two programs are accidentally or maliciously similar, or identify what types of transformations have been applied to a program to hide plagiarism. Thirdly, a useful feature of SCPDTs is the ability to provide a graphical overview of what code fragments are similar or identical between two programs (for example, as provided by MOSS and JPlag). This is used as evidence by a human reviewer investigating if plagiarism has occurred. It would be helpful for BPlag to be extended with such a feature. Finally, there are many similar fields to SCPD, for example code clone detection and program plagiarism detection. It would be interesting to explore if the basic approach of BPlag could be applied to these fields. For example, code clones or program theft could be identified by detecting similar PIDG sub-graphs.

\bibliographystyle{IEEEtran}
\bibliography{references}

\end{document}